\documentclass[twocolumn]{autart}
\usepackage{amsmath}
\usepackage{graphicx}
\usepackage{amssymb}
\usepackage{color}
\usepackage{subcaption}
\usepackage[english]{babel}
\usepackage{tcolorbox}
\newtheorem{lemma}{Lemma}

\def\deff{\stackrel{\triangle}{=}}
\begin{document}

\begin{frontmatter}
\title{Geometric PID Controller for Stabilization of Nonholonomic Mechanical Systems on Lie Groups\thanksref{footnoteinfo}}

\thanks[footnoteinfo]{This paper was not presented at any IFAC
meeting. Corresponding author Ravi N Banavar Tel. 022-25767888.}

\author[Ram]{Rama Seshan}\ead{ ramaseshanc@gmail.com},
\author[Ravi]{Ravi N Banavar}\ead{banavar@iitb.ac.in},
\author[Maithripala]{D H S Maithripala}\ead{smaithri@pdn.ac.lk},
\author[Arun]{Arun D Mahindrakar}\ead{arun\_dm@iitm.ac.in}
\address[Ram]{Graduate student, Department of Electricl Engineering, IIT Madras, Chennai}                                            
\address[Ravi]{Systems \& Control Engineering, IIT Bombay, Mumbai}
\address[Maithripala]{Faculty of Engineering, University of Peradeniya, Peradeniya, Sri Lanka}
\address[Arun]{Department of Electricl Engineering, IIT Madras, Chennai}

\begin{keyword}                           
Geometric methods; PID controller;               
\end{keyword}                             
\begin{abstract}
The PID controller is an elegant and versatile controller for set point tracking in double integrator systems of which mechanical systems evolving on Euclidean space constitute a large class. But since mechanical systems are typically constrained interconnections of rigid bodies whose configuration space is $SE(3)$, which is not even topologically Euclidean, a geometric PID controller has been developed for mechanical systems evolving on Lie groups. In this work, we extend the framework to such systems which have  nonholonomic constraints. It encompasses many practically applicable mechanical systems encountered in robotics as robots are constrained interconnections of rigid bodies where the constraints could either be holonomic or nonholonomic.
\end{abstract}
\end{frontmatter}
\section{Introduction}
In standard control theory, the state of the dynamical system to be controlled is typically a finite dimensional Euclidean space which is described by $\mathbb{R}^n$. But as it is familiar, the state of many mechanical systems evolve on non-Euclidean spaces. As a simple example, the configuration of a rigid body is characterised as an element in the Lie group $SE(3)$. Since robots are constrained interconnections of rigid bodies, the configuration spaces of mechanical systems are submanifolds of Cartesian product of $SE(3)$. The manifold $SE(3)$ is not metrically or even topologically equivalent to the Euclidean space of its same dimension. In such cases, geometric control theory is the apt framework in analysis and control of such systems. Because we do not typically have a single coordinate system covering the entire configuration space, we need coordinate invariant intrinsic geometric techniques for controlling the dynamics in the entire configuration space, which is what geometric control theory provides \cite{Bullo}. In smooth manifolds that are topologically not Euclidean spaces, typically there are topological restrictions that prevent global asymptotic stability by smooth control laws \cite{Bhat}. 

Traditionally, any mechanical system in Euclidean space is a second-order system with the state-space consisting of the configurations and their velocities. We know that for second-order Euclidean systems, a PID controller guarantees global exponential stability even in the presence of constant disturbances.

The PID controller is elegant for second-order systems since it has only three parameters - the proportional (P), derivative (D) and the integral (I) gains - $k_P,k_D,k_I$, which make it practically the most widely used controller. But when the dynamics of a mechanical system evolves on a nonlinear configuration manifold, the design of a PID controller is not straightforward and challenging.

\textit{But intrinsically, all mechanical systems are double integrators in the sense that the acceleration vector is directly related to the force co-vector. Therefore, an intrinsic treatment of the mechanical system is required to apply PID control techniques for mechanical systems evolving on non-Euclidean configuration spaces.}

Traditional PID control design in Euclidean space consists of the following steps:
\begin{enumerate}
    \item  Defining the tracking error as the difference between the current configuration vector and the desired configuration vector.
    \item Deriving the error dynamics, which expresses the time evolution of error.
    \item Driving the error to zero with three control forces, proportional to the error, the time-integral of the error, and the time-derivative of the error respectively.
\end{enumerate}
Geometric control methods have been developed for a wide variety of systems. There have been works that generalize PID controller to very specific class of mechanical systems without any constraints. For example, \cite{Madhu} generalizes it for hoop robots on an inclined plane. In \cite{Somasiri} and \cite{Lee}, it is generalized for the quadrotor and UAV on SE(3) respectively.  Intrinsic observers and filters have also been considered by the authors in \cite{MDB} and \cite{Mahony}.

 So, to extend PID controller to arbitrary mechanical system evolving on a Lie group like direct product of $SE(3)$, we need to generalize these notions from the Euclidean case. The first of these is dealt by using the gradient of a globally defined smooth function whose gradient mimics the properties of the error vector in the Euclidean case. Such functions are called polar Morse functions whose existence has been proven in any smooth manifold \cite{Morse} and for smooth compact manifolds with boundary \cite{Kodit}. The derivative term has been generalized by using the left translation map to bring the reference velocity vector to the same tangent space as the actual velocity vector \cite{Bullo}. The parallel transport map also can be used to do so but it is not intrinsic, and its computation is complicated and requires commitment to a particular choice of coordinates which makes its applicability limited to only guaranteeing local stability. Finally, using PD control, almost-global stability of a desired constant configuration on a Riemannian manifold  has been further developed in \cite{Bullo}, \cite{BulloMurray}, \cite{BulloMurray1}, \cite{Leee}, \cite{MBD}, \cite{MDB}. Finally the integral action was generalized and the PID controller was established to be almost globally asymptotically stable in the presence of constant disturbances in \cite{DHSM}. Integral control has also been utilized in \cite{Zhang}. 

 In this work, we focus on extending the results of the unconstrained PID controller developed in \cite{DHSM} to nonholonomically constrained mechanical system on Lie groups which are typically Cartesian products of SE(3). While this extension is mathematically non-trivial and poses a theoretical challenge, it is also practically relevant since most mechanical systems are not just interconnections of rigid bodies but are constrained interconnections of rigid bodies and hence wide real-life examples are covered which include but not limited to the mobile robot, Segway, spherical robots, spherical pendulum and car-like robots. Further, even if the constraints are holonomic, (the distribution defining the constraint is integrable), the integral manifold in which the states evolve typically is not a Lie group as all submanifolds of a Lie group are not Lie groups. So, even in the holonomically constrained case, the system cannot be treated as an unconstrained system evolving on a Lie group but it can only be treated as evolving in the distribution only as treated in the present work. Also, in nonholonomic system, it is not possible to actuate the system in the direction of the constraints and any attempt to do so will be cancelled by the constraint forces. So such systems are classified as underactuated and the fully actuated PID controller fails for such systems. But in the framework developed in this paper, such systems are considered fully actuated when the actuating forces are just able to span the codistribution and are hence treated on par with fully actuated unconstrained mechanical systems. So this framework of velocities of the system evolving in a constraint distribution, has wide applicability.
 
 \section{Familiar results from elementary mechanics and control theory in Euclidean space}
\subsection{A simple case study in elementary mechanics}
Consider a point particle in a plane that is constrained to move in a circular groove of radius $r$ as shown below, and acted upon by an external force $\vec{F}$. We know from elementary high school mechanics that there will be a normal force exerted by the constraining groove on the mass so that it always stays in the circle. 

\begin{figure}[h]
	\center{\includegraphics[scale=0.5]{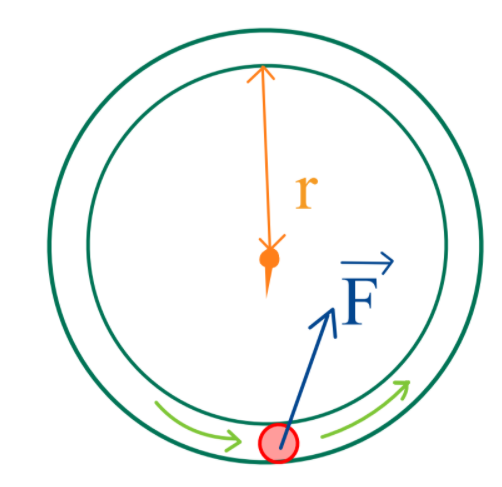}}
	\caption{Simple constrained mechanical system on the plane}
	\label{fig:mylabel1}
\end{figure}

At a particular point, let $D$ and $D_{\perp}$ be the tangent and normal directions. Note here that the tangent and normal directions vary from point to point so that the direction of the normal is changing depending on where the particle is. 

\begin{figure}[h]
	\center{\includegraphics[scale=0.5]{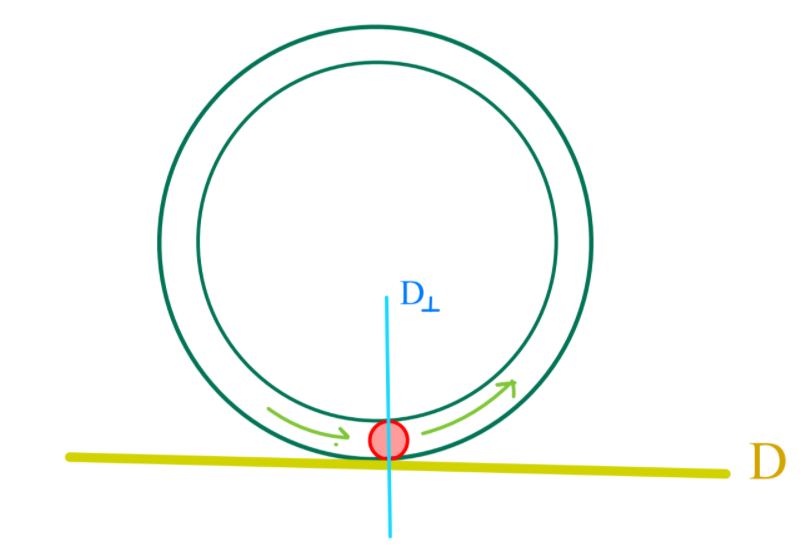}}
	\caption{Tangent and normal directions $D,D_{\perp}$}
	\label{fig:mylabel1}
\end{figure}

We see that the groove constrains the velocity of the particle to stay in the tangent direction $D$ at every point. 

Now, let us split the Newton's equations of motion into two parts - the tangential and normal directions. Let $\vec{a}_D,\vec{a}_{D_{\perp}}$ be the tangential and normal component of the acceleration. Let $\vec{F}_D,\vec{F}_{D_{\perp}}$ be the components of the applied force in the tangent and normal direction respectively. 

Now, let us see what the equations of motion are, in the tangential and normal direction. In the tangential direction, we have as usual that mass times the tangential acceleration equals the force in the tangential direction and hence
\begin{align}
m\vec{a}_{D}=\vec{F}_D
 \label{eqntang}
\end{align}

We also know from high school analysis that in the normal direction, there is a constraint force $\vec{N}$  exerted by the surface on the mass to enforce the constraint. 

\begin{figure}[h]
	\center{\includegraphics[scale=0.7]{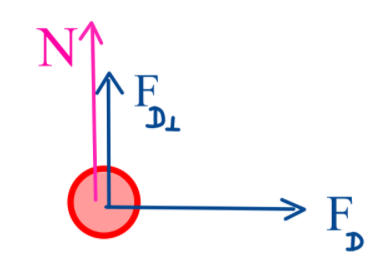}}
	\caption{Forces in the Tangent and normal directions $D,D_{\perp}$}
	\label{fig:mylabel1}
\end{figure}

\begin{align}
m\vec{a}_{D_{\perp}}=\vec{F}_{D_{\perp}}+\vec{N}
 \label{eqnnl}
\end{align}

If the constraint direction $D$ is same at every point, (like if it is constrained to move in the horizontal direction $D$ forever instead of in a circle), then it is same as requiring that the normal component of acceleration $\vec{a}_{D_{\perp}}$ vanish and hence we should have that the constraint force $\vec{N}$ cancels the applied force in the normal direction and hence
\begin{align}
\vec{N}=-\vec{F}_{D_{\perp}}
 \label{eqnforcen}
\end{align}

But the constraint direction $D$ is not constant and changes from point to point. So, we cannot have that the constraint force just cancels the applied force in the normal direction but also helps the particle to turn around so that the particle will have its velocity pointed in the correct direction when it goes to a nearby point a while later. In case of a constraint being a circle of radius $r$, we precisely know that this is the centripetal force $\vec{F}_{C}=mr\dot{\theta}^2$ where $\theta$ is the angular coordinate. This term is due to the constraint direction $D$ changing from point to point, and hence the particle in addition to not having normal velocity right now, has to turn in the right direction so that at the next instant, its direction is correctly oriented along the constraint direction at that point and this requires a push in the normal direction. So we have that

\begin{align}
\vec{N}=-\vec{F}_{D_{\perp}}+\vec{F}_C
 \label{eqnforcen}
\end{align}

So \textbf{the normal constraint force has two parts - one to cancel the applied external force in the normal direction and another one to turn the particle appropriately so that it is correctly oriented along a new direction in a nearby point, after it moves. This term is the centripetal force for the case of a particle moving in a circle and it extends for other surface also by fitting the best approximating circle that fits the surface around a point}.

We know that due to the constraint, the velocity of the particle $\vec{v}$ can have only tangential component and no normal component at all times. i.e. $\vec{v}(t)=\vec{v}_D(t)$. So, the normal component of the velocity, if initially vanishes, has to vanish for all time. i.e
\begin{align}
\vec{v}_{D_{\perp}}(0)=\vec{0}\Longrightarrow \vec{v}_{D_{\perp}}(t)=\vec{0}\hspace{2mm}\forall t\\
\Longrightarrow \frac{d}{dt}\vec{v}_{D_{\perp}}(t)=0
 \label{eqnexample}
\end{align}

But what we have in LHS of the equation of motion in the normal direction is the normal component of acceleration which is $\vec{a}_{D_{\perp}}=\bigg(\frac{d\vec{v}}{dt}\bigg)_{D_{\perp}}$ (the order of projection and derivative are reversed between the normal acceleration and derivative of normal component of velocity)

If $P_{D_{\perp}}$ denotes the perpendicular projection into the direction $D_{\perp}$, then by the chain rule and Liebnitz rule for calculus, we have
\begin{align*}
    \vec{a}_{D_{\perp}}=P_{D_{\perp}}\bigg(\frac{d\vec{v}}{dt}\bigg)=\frac{d}{dt}\bigg(P_{D_{\perp}}\vec{v}\bigg)-\bigg(\frac{d}{dt}P_{D_{\perp}}\bigg)\vec{v}
\end{align*}

Now, enforcing the constraint is equivalent to making $\frac{d}{dt}\bigg(P_{D_{\perp}}\vec{v}\bigg)=0$ and hence putting this for $\vec{a}_{D_{\perp}}$ in Newton's equation, we get
\begin{align}
\vec{N}=-\vec{F}_{D_{\perp}}-\bigg(\frac{d}{dt}P_{D_{\perp}}\bigg)\vec{v}
\label{compare}
\end{align}
So we realise that comparing the equations \ref{eqnforcen} and \ref{compare}, we get the centripetal term as
\begin{align}
\vec{F}_C=-\bigg(\frac{d}{dt}P_{D_{\perp}}\bigg)\vec{v}
\label{cpt}
\end{align}
So this extra centripetal term arises because $D$ changes from point to point and hence the map $P_{D_{\perp}}$ also changes in the direction of $\vec{v}$ and hence it vanishes only if $P_{D_{\perp}}$ is constant - i.e. the velocity constraint subspace direction at neighbouring locations agree with the velocity constraint subspace at the given point. So, we see that even in constrained Euclidean systems, the equations of motion split in the tangential and normal direction. The equation of motion in the  tangential direction is nothing but the projection of intrinsic Newton's equations whereas in the normal direction, the equation of motion prescribes the constraint force offered by the surface to enforce the constraint.

\subsection{Classical PID control - a geometric review}
Consider a dynamical system evolving in Euclidean space $\mathbb{R}^n$, that is supposed to be stabilised at the point $\vec{x}_d$. Now, the central ingredients in PID control are that of error vector $\vec{e}=\vec{x}-\vec{x}_d$, derivative of error vector $\vec{\dot{e}}$ and integral error vector $\vec{z}=\int \vec{e} dt$. 

This notion of error vector arises naturally since $\mathbb{R}^n$ is a vector space, which is equipped with an inner product. The Euclidean metric on $\mathbb{R}^n$ is the guide to defining the error vector $\vec{e}=\vec{x}-\vec{x}_d$. The error vector is nothing but the gradient of the metric function as

\begin{align}
\vec{e}=\vec{x}-\vec{x}_d=\nabla V
\label{euclideanerror}
\end{align}
where $V$ is the metric function
\begin{align}
V(\vec{x})=\frac{1}{2}||\vec{x}-\vec{x}_d||^2
\label{eucideandistance}
\end{align}

The metric function $V(\vec{x})$ in $\mathbb{R}^n$ has some nice properties that enable it to serve as a guide for giving the error direction at each point - they are 
\begin{itemize}
    \item It is zero and minimum only at the desired point $\vec{x}_d$ and positive everywhere else - so the problem of making $\vec{e}=0$ reduces to making the scalar function $V(\vec{x})$ minimum - which is zero
    \item Since the gradient points along the direction of maximum increase of $V$ and hence the negative gradient pointing towards the direction of maximum decrease of $V$, its negative gradient serves as a candidate in prescribing the direction in which one has to move to reduce it to a minimum since it is the direction in which the best possible reduction occurs
    \item Since it has no local minima or even local extrema anywhere else other than the desired point, the gradient never vanishes other than at the desired point and hence as long as we are away from the desired point, we will always move away from it and towards $\vec{x}_d$, to reduce $V$ further
\end{itemize}
So we now see that the error vector is not that special as far as its use in PID control is concerned - the gradient of any suitable function $V$ satisfying the above properties can serve as the error vector as well and the stability and convergence of PID control rests only on the above properties. These are what will be carried forward in the general setting of the paper.

Now, the derivative of the error is simply the time derivative of the error vector - the componentwise differentiation in Euclidean space and the integral error $\vec{z}$ can also be defined equivalently by the dynamical system
\begin{align}
\dot{\vec{z}}=\vec{e}
\label{inteuc}
\end{align}

where $\vec{z}$ is the integral error. 

For the second order mechanical system in Euclidean space defined by

\begin{align}
m\ddot{\vec{x}}=\vec{F}
\label{simplemech}
\end{align}
,the PID controller is nothing  but the force law
\begin{align}
\vec{F}=-k_P \vec{e}-k_D\vec{\dot{e}}-k_I \vec{z}\\
\dot{\vec{z}}=\vec{e}
\label{simplemechforcelaw}
\end{align}

The integral error term is needed when the system has to deal with other constant disturbance forces apart from the ones accounted for. Let us call this $\vec{D}$. Rewriting the dynamical system with the disturbance, we get

\begin{align}
m\ddot{\vec{x}}=\vec{D}+-k_P \vec{e}-k_D\vec{\dot{e}}-k_I \vec{z}\\
\dot{\vec{z}}=\vec{e}
\label{simplemechfbackdyn}
\end{align}

Let us establish the stability result for this Euclidean system so that one can appreciate it for the general case of the paper.

First, the equilibria $(\vec{x}_{eq},\vec{z}_{eq})$ for this dynamical system is obtained by setting $\dot{x}=\ddot{x}=\dot{z}=\vec{0}$. Setting $\dot{z}=0$ implies $\vec{e}=\vec{0}$ an hence we have that any equilibrium for this system should satisfy $\vec{e}=0,\dot{x}=0$ and $\vec{z}_{EQ}$ will depend on the disturbance $\vec{D}$. If $\vec{D}=0$, then $\vec{z}_{EQ}=0$. In this case, we shall prove global exponential stability.

Consider the Lyapunov function $W$ for the dynamical system in $(\vec{x},\vec{\dot{x}},\vec{z})$ defined as
\begin{align}
W=k_P\frac{1}{2}||\vec{e}||^2+\frac{1}{2}||\vec{\dot{e}}||^2+\frac{\gamma}{2}||\vec{z}||^2+\alpha \vec{e}\cdot \vec{\dot{e}}+\beta \vec{z}\cdot \vec{\dot{e}}+\sigma \vec{e}\cdot \vec{z}
\label{Lyapunov}
\end{align}
where $\gamma>0,\alpha,\beta,\gamma$ are constants to be determined in due course.

In matrix form, with the product of vectors being understood as dot product or inner product, this function $W$ can be written as
\begin{align}
W=\frac{1}{2}\begin{bmatrix}
\vec{e} & \vec{\dot{e}} & \vec{z} 
\end{bmatrix}\begin{bmatrix}
k_P & \alpha & \sigma \\ \alpha & 1 & \beta \\ \sigma & \beta  & \gamma
\end{bmatrix}\begin{bmatrix} \vec{e} \\ \vec{\dot{e}} \\ \vec{z}
\end{bmatrix}=\frac{1}{2}u^TPu
\label{Lyapunovmatrixform}
\end{align}
where $u=\begin{bmatrix} \vec{e} & \vec{\dot{e}} & \vec{z}
\end{bmatrix}^T$ and $P$ denoting the square matrix.

First, we see that this function is positive definite about the only equilibrium $(\vec{e},\vec{\dot{e}},\vec{z})=(0,0,0)$ if the square matrix in the defining equation \ref{Lyapunovmatrixform} is positive definite which boils down to the conditions presented in the lemma below that follows from matrix analysis.
\begin{lemma}
    \label{posdeflemma}
    The function $W$ or matrix $P$ is positive definite if and only if the following conditions below hold:
    \begin{align}
\gamma>\beta^2\\
k_p>\frac{\gamma \alpha^2+\sigma^2+2\alpha\sigma\beta}{\gamma-\beta^2}
\label{posdefcond}
\end{align}
\end{lemma}
Let us now evaluate the derivative of $W$, along the trajectories of the dynamical system evaluated as 
\begin{align}
\dot{W}=\begin{bmatrix}
\vec{e} & \vec{\dot{e}} & \vec{z} 
\end{bmatrix}\begin{bmatrix}
k_P & \alpha & \sigma \\ \alpha & 1 & \beta \\ \sigma & \beta  & \gamma
\end{bmatrix}\begin{bmatrix} \vec{\dot{e}} \\ \vec{\ddot{e}} \\ \vec{\dot{z}}
\end{bmatrix}
\label{Wdot}
\end{align}
which on simplification becomes
\begin{align}
\dot{W}=&k_P \vec{e}\cdot \dot{\vec{e}} +\alpha(\vec{e}\cdot \ddot{\vec{e}}+\dot{\vec{e}}\cdot \dot{\vec{e}})+\sigma(\vec{e}\cdot \dot{\vec{z}}+\vec{z}\cdot \dot{\vec{e}})+\dot{\vec{e}}\cdot \ddot{\vec{e}}+\nonumber\\&\beta(\dot{\vec{e}}\cdot \dot{\vec{z}}+\vec{z}\cdot\ddot{\vec{e}})+\gamma(\vec{z}\cdot\dot{\vec{z}})
\label{Wdotcontnd}
\end{align}
Putting for $\ddot{e},\dot{z}$ from the dynamic equation \ref{simplemechfbackdyn}, we get $\dot{W}$ expression as another quadratic form in $(\vec{e},\dot{\vec{e}},\vec{z})$ that is given below

\begin{lemma}
	\label{Wdotlemmaeuc}
	\begin{align}
-2\dot{W}&=\nonumber\\&\begin{bmatrix}
\vec{e} & \vec{\dot{e}} & \vec{z} 
\end{bmatrix}\begin{bmatrix}
2(k_P\alpha-\sigma) & k_D\alpha-\beta & k_I\alpha+k_P\beta-\gamma \\ k_D\alpha-\beta & 2(k_D-\alpha) & k_I+\beta k_D -\sigma \\ k_I\alpha+k_P\beta-\gamma & k_I+\beta k_D -\sigma  & 2\beta k_I
\end{bmatrix}\begin{bmatrix} \vec{\dot{e}} \\ \vec{\ddot{e}} \\ \vec{\dot{z}}
\end{bmatrix}\nonumber\\&=u^TQu
\label{Wdot}
\end{align}
\end{lemma}

Now, the function $\dot{W}$ is negative definite (required for applying Lyapunov-LaSalle Theorems) if and only if the matrix $Q$ is positive definite. 

So the design problem now reduces to choosing $\alpha,\beta,\gamma$ that ensure that in some ranges of $k_P,k_I,k_D$, the matrices $P,Q$ are positive definite 

Now choosing the following $\alpha,\beta,\gamma$ as
\begin{align}
\alpha&=\frac{k_I}{k_D^2}\\
\beta&=\frac{k_I}{k_D}\\
\gamma&=\frac{k_I^2+k_Ik_Pk_D}{k_D^2}
\label{designofW}
\end{align}
we see that it obviously satisfies $\gamma>\beta^2$ (one of the conditions in Lemma 1 for positive definiteness of $P$). Freezing this as $\alpha,\beta,\gamma$, we get the matrix $Q$ as 
\begin{align}
\begin{bmatrix}
2(\frac{k_Pk_I}{k_D^2}-\sigma) & 0& 0  \\ 0 & 2(k_D-\frac{k_I}{k_D^2}) & 2k_I-\sigma \\ 0&  2k_I-\sigma   & 2\frac{k_I^2}{k_D}
\end{bmatrix}
\label{Qmatrixfreeze}
\end{align}

Now for $Q$ to be positive definite, we must have

\begin{lemma}
    \label{posdefQlemma}
    The matrix $Q$ is positive definite if and only if the following conditions below hold:
    \begin{align}
\sigma<\frac{k_Pk_I}{k_D^2}\\
4\bigg(k_D-\frac{k_I}{k_D^2}\bigg)\frac{k_I^2}{k_D}>(2k_I-\sigma)^2
\label{posdefcondQ}
\end{align}
\end{lemma}

Now, making the following substitutions $\sigma=2Kk_I$ and $\delta=K-1$, we have the conditions as
\begin{align}
k_P>2Kk_D^2\\
k_D^3>\frac{k_I}{1-\delta^2}
\label{posdefcondQ2}
\end{align}
Note that from 
Note that for $P$ to be positive definite, from Lemma \ref{posdeflemma}, the condition Equation \ref{posdefcond} also requires $k_P>\frac{\gamma\alpha^2+\sigma^2+2\alpha\sigma\beta}{\gamma-\beta^2}$. Plugging the definitions of $\sigma$ and expressions for $\alpha,\beta,\gamma$ from Eqn \ref{designofW}, combining everything, we have
 \begin{align}
 \label{kpcondn}
k_P^2>\frac{k_I^3+4K^2k_Ik_D^6+4Kk_I^2k_D^3}{k_D^5}
\end{align}
The choice of gains is as follows. We first choose any $k_I>0$. We then select
 any $K$ (which in turn fixes $\sigma$ as we have $\sigma=2Kk_I$) such that $0<K<2$ (which implies $0<\delta<1$), which implies $1-\delta^2>0$. With this choice, we choose $k_D$ and $k_P$ to satisfy the conditions needed for positive definiteness of $P$ and $Q$ and hence we finally have
\begin{thm}

    \label{mainthminrn}
  The system is stable and converges to the equilibrium for the following choices of $k_P,k_I,k_D$ (by Lyapunov-LaSalle Theorems) 
  
    \begin{align}
    k_I>0\\
    k_D^3>\frac{k_I}{1-\delta^2}\\
    k_P>\max \bigg(2Kk_D^2,\sqrt{\frac{k_I^3+4K^2k_Ik_D^6+4Kk_I^2k_D^3}{k_D^5}}\bigg)
\end{align}
\end{thm}

In the case of a constant disturbance, the equilibrium is no longer $(\vec{e},\vec{\dot{e}},\vec{z})=(0,0,0)$ but it is now $(0,0,\frac{1}{k_I}\vec{D})$ and by defining $\vec{z}'=\vec{z}-\frac{1}{k_I}\vec{D}$, we can exactly do the same Lyapunov analysis, replacing $\vec{z}$ by $\vec{z}'$ (which does not affect the integrator dynamics as $\vec{\dot{z}}=\vec{\dot{z}}'=\vec{e}$. So the system still converges to $\vec{e}=\vec{\dot{e}}=0$ at equilibrium but with a different integrator error value.

In the generic case, the algebra will be exactly similar and only the interpretation will be different. The configuration space will no longer be Euclidean and all that we need to do is find a clever way around by finding a substitute for the derivative of vectors (which will turn out to be the covariant derivative $\nabla$) and for the error function (which will turn to be called as the polar Morse function). Once that is done, the design is exactly same as the Euclidean case only with a minor modification of $K$ that we will reserve to analyze for the general case. Also additionally, we will have to deal with designing the controller in the case when the system is having a non-holonomic constraint on velocity. The algebra of design of $k_P,k_I,k_D$ is exactly whatever was done above. 

\subsection{PID with Feedforward Control for time varying trajectories - a review}
Consider a system as usual given by the unit mass Newtonian double integrator $\ddot{\vec{x}}=\vec{F}$. Now, let us have a time varying reference trajectory $\vec{x}_r(t)$ which is no longer a set point. Then it is a well known technique to convert this into a constant set point problem. 

Then we can define the tracking error as $\vec{e}(t)=\vec{x}(t)-\vec{x}_r(t)$. But since the reference is time varying, its first and second derivatives no longer vanish. So we have $\dot{\vec{e}}(t)=\dot{\vec{x}}(t)-\dot{\vec{x_r}}(t)$,$\ddot{\vec{e}}(t)=\ddot{\vec{x}}(t)-\ddot{\vec{x_r}}(t)$. So, expressing the system dynamics in terms of $\vec{e}$, we get
\begin{align}
    \label{edynrn}
    \ddot{\vec{e}}(t)+\ddot{\vec{x_r}}(t)=\vec{F}
\end{align}
Now, defining $\vec{U}=\vec{F}-\ddot{\vec{x_r}}(t)$, we have
\begin{align}
    \label{edynrnu}
    \ddot{\vec{e}}(t)=\vec{U}
\end{align}
So, now the error dynamics of the system behaves exactly as the original double integrator system except that it has input $\vec{U}$ instead of $\vec{F}$. So, design $\vec{U}$ by the Lyapunov technique in the previous subsection and then apply input $\vec{F}=\vec{U}+\ddot{\vec{x_r}}(t)$. The extra term to be added to $\vec{F}$ which is the acceleration of the reference trajectory is called the feedforward term will also find its analog. 

To convert a tracking problem into a regulation problem about $\vec{0}$, we defined error by subtracting the actual trajectory from the reference trajectory which is a legitimate operation in Euclidean space. But for arbitrary spaces, this is impossible unless there is some operation like vector subtraction. It turns out that a Lie group with a group operation is the ideal candidate for such generalization of vector subtraction.

 \section{Mathematical Preliminaries and Notation}
 We recall in this section, the mathematical preliminaries on Lie groups, Riemannian geometry and nonholonomic mechanical systems. The notations used are standard as in \cite{Bullo}, \cite{Frankel}, \cite{Marsden}.
 
 We denote by $G$, a Lie group with binary operation '$.$'. Its identity element is denoted by  $e$ and its Lie algebra (tangent space at identity) is denoted by $\mathfrak{g}$. We denote by $T_gG$, the tangent space of the Lie group $G$ at the point $g$ and by $TG$, the tangent bundle of $G$. We denote the dual space of a vector space by putting a $*$ in the superscript. For example, $T_{g}^{*}G$ stands for the dual of $T_gG$ which is the definition of the cotangent space of $G$ and we denote by $T^*G$, the cotangent bundle of $G$. Let $L_g$ denote the left translation map in $G$ that translates every element in $G$ by $g$ from the left. The derivatives of any smooth map between two manifolds $f:M\rightarrow N$ are as usual denoted by $f_*:T_mM\rightarrow T_{f(m)}N$ and the pull back map is denoted by $f^*:T^*_{f(m)}N\rightarrow T^*_mM$.  
 
  We assume throughout that $G$ is endowed with a left invariant metric tensor denoted by $\mathbb{I}$. Unlike standard Riemannian geometry, we allow for the possibility of $\mathbb{I}$ to be singular. That is, $\mathbb{I}$ is required to be only positive semi-definite and not positive definite everywhere as is required in standard Riemannian geometry. This is motivated by the fact that several important practical mechanical systems (like the spherical pendulum for instance) has an inertia matrix that is singular. This will have several consequences. Firstly, the canonical map from the tangent space $T_gG$ to the cotangent space $T_g^*G$ defined by $v \in T_gG \rightarrow  \omega (v) \in T_g^*G$ defined by $\omega(v) (x)=\mathbb{I}(v,x)$ for any $x \in T_gG$ is still well-defined. Thus we may associate a covector canonically to every vector by freezing one of the slots of the metric with the given vector. But what is now lost is the invertibility of this map. We can no longer associate a vector to call covectors uniquely. Thus equations or expressions on the cotangent space of the manifold $G$ cannot be interpreted in tangent space by inverting the metric tensor map $\mathbb{I}$. So we have to deal with equations in the cotangent space as such.

Denoting by the letter $V$, a member of a special class of smooth real valued function on $G$, called the polar Morse function that will be elaborated in the text. Denote by $C^{\infty}(G)$, the algebra of all smooth real-valued functions on $G$. A tangent vector at a given point $g$, as is a derivation on $C^{\infty}(G)$ - that is a map that associates to each function $f\in C^{\infty}(G)$, a real number called the directional derivative operator, that obeys linearity and Liebnitz rule. A vector field in $G$, denoted by letters like $X,Y$ as one might recall from standard differential geometry is a smooth assignment of tangent vectors in the tangent space at every point in $G$ and similarly a covector field, of cotangent vectors, and a tensor field, of tensors. For example, the metric is a $(0,2)$ tensor field on $G$.  The Lie Bracket is a binary operation on the space of vector fields defined by $[X,Y](f)=X(Y(f))-Y(X(f))$ for $f\in C^{\infty}(G)$. It satisfies all the requirements to be a Lie algebra on the space of vector fields.

Next, we denote by $D$, a smooth distribution in $TG$ which is a smooth assignment of a subspace of $T_gG$ to every point in $g\in G$.
Distributions arise in mechanics when there are nonholonomic constraints on the mechanical system - a constraint restricting the possible velocities allowed to be taken by the system that cannot be integrated to yield a constraint on position. The projections of a vector on to these various vector spaces - $D,D^*,D_{\perp},D^*_{\perp}$ are denoted by $P_D,P_{D^*},P_{D_{\perp}},P_{D_{\perp}^*}$.  Since $D$ is smooth, these projections may be regarded as smooth $(1,1)$ tensor fields on $G$. A distribution $D$ is said to closed Lie algebraically if the Lie Bracket of every pair of vector fields in $D$ lies within $D$ itself. We also denote by $\mathfrak{D}$, the Lie algebraic closure of the given distribution $D$, also referred to as the minimal Lie algebra containing $D$. This is the smallest Lie algebraically closed distribution that contains $D$. It is imperative to recall Frobenius theorem in distributions that states that a distribution $D$ is integrable if and only if $D=\mathfrak{D}$. We denote by $D^*$, the image of the distribution, under the metric map, that is, $D^*:=\mathbb{I}(D)$. We denote by $D_{\perp}$, the orthogonal complement of $D$ in $TG$ and its image in the cotangent space by $D^*_{\perp}$. They are visualized in Figure \ref{fig:mylabel1}.
\begin{figure}[h]
	\center{\includegraphics[scale=0.4]{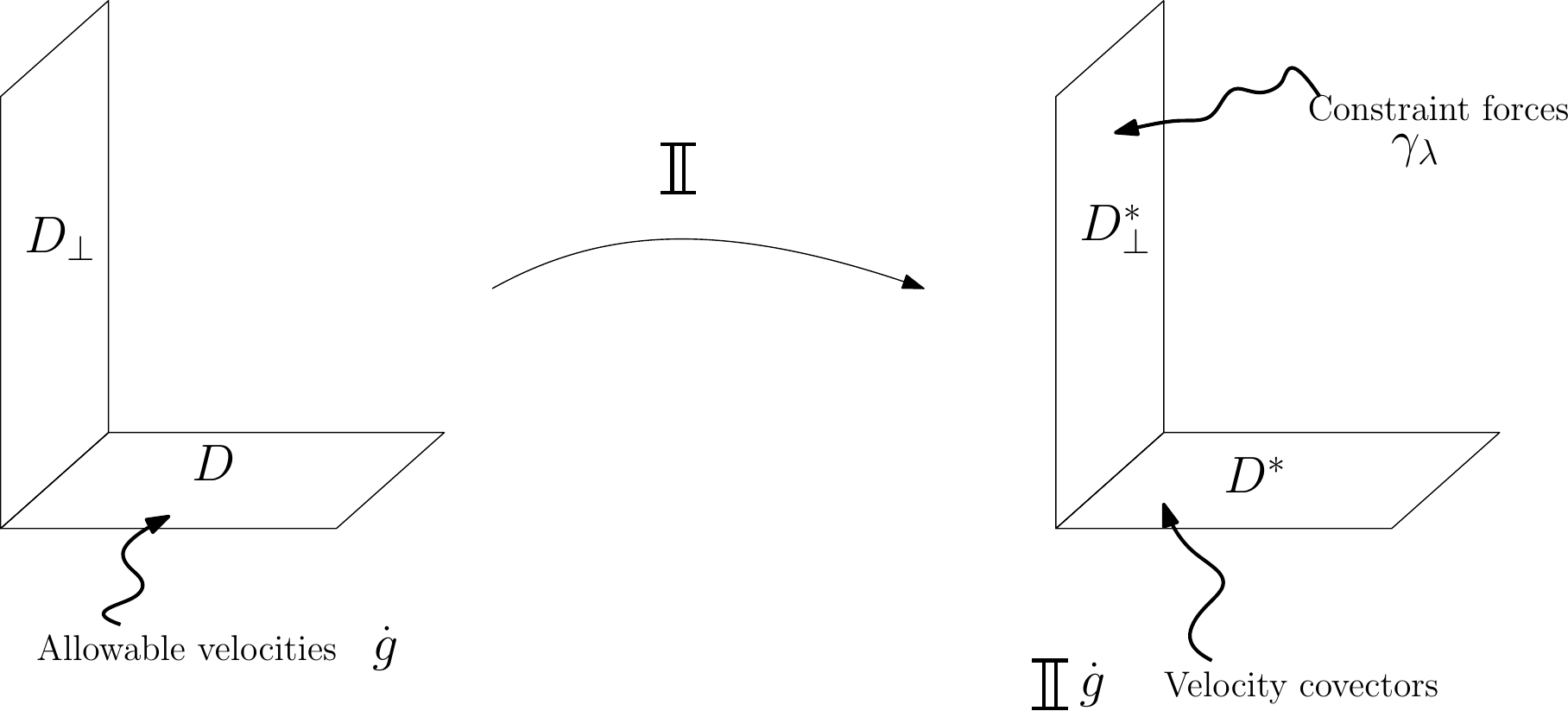}}
	\caption{The constraint subspaces -  $TG=D\oplus D_{\perp}$ and $T^*G=D^*\oplus D^*_{\perp}$ }
	\label{fig:mylabel1}
\end{figure}
The covariant derivative $\nabla$ or an affine connection is a notion of differentiating a tensor field with respect to a vector field to yield another tensor field that is to be regarded as a ``derivative" of the original tensor field in the direction of the given vector field which obeys linearity and Liebnitz rule for  the incoming tensor. A formal definition can be found in  \cite{Morse} . A covariant derivative is said to be metric compatible with a metric $\mathbb{I}$ if the covariant derivative is symmetric and the covariant derivative of the metric tensor with respect to any vector field vanishes. That is $\nabla_X \mathbb{I}=0$ for any vector field $X$. So, this ensures that the metric tensor $\mathbb{I}$ can be taken inside and outside of any covariant derivative freely and this will be the case in all the analysis that follows. In case of non-degenerate metrics found in standard Riemannian geometry, the covariant derivative compatible with a metric is unique but it is not the case so for the degenerate case that we have allowed for here.

With this, we proceed to the PID controller. A PID controller for a regulator problem for dynamical system evolving on $\mathbb{R}^n$, viewed as a Lie group with the operation of vector addition $+$ and identity as $\pmb{0}$, has four central ingredients that is recalled. First is the notion of an error metric which is defined as square of Euclidean distance from the set point (assumed as $\pmb{0}$) and the given point $\pmb{E}$ that is given by $||\pmb{E}||^2$. It gradient of the error function which is $\frac{1}{2}||\pmb{E}||^2$, is what we interpret as an error vector $\pmb{E}$ - the direction that we have to move from $\pmb{E}$ so that the error function increases the maximum. The reason why the error function is suitable for regulator problems is that it satisfies desirable properties - it is minimum in the desired set point $\pmb{0}$ and has no local extremum anywhere else and this ensures that if we follow the negative gradient of the error vector, we reach the global minimum, which is $\pmb{0}$. These motivate the need for defining polar Morse functions. Next, we have the notions of error derivative or error velocity and the notion of an integral error. So, to generalize the PID controller in an arbitrary Lie group with a metric, one has to generalize the notion of the error function, the error vector, the error velocity and the integral error, while respecting the nonholonomic constraints on $G$.

Let $L_g$ denote the left translation by $g\in G$ and $R_g$ denote the right translation by $g \in G$ maps respectively. For any tangent vector $X_g \in T_gG$, we denote by $h.X_g$, the derivative of the left translation map $L_h$ acting on $X$, that is, $h\cdot X_g:=(L_h)_*X_g \in T_{hg}G$ and similarly the notation $X_g\cdot h$ for $(R_h)_*X_g \in T_{gh}G$. Also denote the adjoint map $Ad:T_eG\rightarrow T_eG$ as $Ad_g(X)=g\cdot X \cdot g^{-1}$.

\section{Nonholonomic Mechanical Systems on Lie Groups}
Let $G$ be a Lie group and the configuration space of a mechanical system with a nonholonomic constraint distribution $D$ (assumed to be smooth). Let $\mathbb{I}$ be the left invariant metric on $G$ (possibly degenerate). So, a constrained mechanical system is denoted by the quartet $(G,\cdot,\mathbb{I},D)$. We make the following assumptions:
\begin{itemize}
	\item $\mathbb{I}$ restricted to $D$ is injective (non-degenerate) which is equivalent to existence of left invertibility of $\mathbb{I}$.
	\item $D$ is of constant dimension and smooth.
\end{itemize}
Let $D^*=\mathbb{I}(D)$ be the dual image of D in the cotangent space and $D_{\perp}^*$ be its perpendicular subspace. Let $P_{D^*}$ and $P_{D_{\perp}^*}$ be the corresponding projection maps onto the respective subspaces. Note that they are smooth (1,1) tensor fields on G. The subspaces are visualized as shown in Figure \ref{fig:mylabel1}.
%
%
\begin{rem}
	 The assumption of $\mathbb{I}$ being injective when restricted to $D$ ensures that if the constraint is satisfied in cotangent space (or more precisely, if the cotangent vector lies in the image of the distribution under $\mathbb{I}$), it is satisfied in tangent space as well. This is easily seen because the assumption of $\mathbb{I}$ being injective implies that it is invertible as a map $\mathbb{I}:D\rightarrow \mathbb{I}(D)=D^*$. So, $\mathbb{I}\dot{g}\in D^* \Longrightarrow \dot{g}\in D$ due to left invertibility.
\end{rem}

Now, if $g(t)$ is the trajectory of an unconstrained mechanical system $(G,\cdot,\mathbb{I})$, the intrinsic Newton's equations on G for the unconstrained system reads as:
\begin{align}
\mathbb{I}\nabla_{\dot{g}}\dot{g} = \gamma \label{eqnintrinsic}
\end{align}
where $\gamma$ represents the external force 1-forms.

In the case of constrained systems, by the well-known  Lagrange-d'Alembert principle that there is a constraint force $\gamma_{\lambda}$ acting in $D^*_{\perp}$ to enforce the constraint on the trajectories of the mechanical system. We know that the applied force covector can influence the dynamics only along the constraint codistribution while the constraint force acting on the complementary space of the codistribution ensures that the trajectory satisfies the constraint. (that is, its velocity points along the distribution at all times if it starts on the distribution initially). So, projecting the intrinsic Newtonian dynamics along $D^*$ and $D_{\perp}^*$, and observing that the constraint force can act only on $D^*_{\perp}, $we get
\begin{align}
P_{D^*}(\mathbb{I}\nabla_{\dot{g}}\dot{g})=P_{D^*}\gamma \label{eqnmechD} \\
P_{D_{\perp}^*}(\mathbb{I}\nabla_{\dot{g}}\dot{g})=P_{D_{\perp}^*}\gamma+\gamma_{\lambda}. \label{eqnmechDperp}
\end{align}
Now, since the covariant derivative obeys Leibnitz rule, we have
\begin{align}
\nabla_{X}(P_{D_{\perp}^*}\mathbb{I}Y)=(\nabla_{X}P_{D^*_{\perp}})\mathbb{I}Y + P_{D^*_{\perp}}(\nabla_{X}\mathbb{I}Y), \label{Liebnitz}
\end{align}
where $(\nabla_{X}P_{D^*_{\perp}})$ is the covariant derivative of $P_{D_{\perp}^*}$ when regarded as a tensor field in $G$. Note that as the
metric tensor is itself invariant (we have a metric compatible connection), we have $\nabla_{(X)}\mathbb{I}=0$ and hence $\mathbb{I}$ can be freely pulled in and out of the covariant derivative.
Using Leibnitz rule in the second equation of Newton's dynamics  \eqref{eqnmechDperp}, gives
\begin{align}
P_{D^*}(\mathbb{I}\nabla_{\dot{g}}\dot{g})&=P_{D^*}\gamma \label{LiebD} \\
\nabla_{\dot{g}}(P_{D_{\perp}^*}\mathbb{I}\dot{g})-(\nabla_{\dot{g}}P_{D^*_{\perp}})\mathbb{I}\dot{g}&=P_{D_{\perp}^*}\gamma+\gamma_{\lambda}. \label{LiebDperp}
\end{align}
We know from mechanics that the role of the constraint force $\gamma_{\lambda}$ is to ensure that the velocity of the trajectory stays in D, once it starts in D, or
\begin{align*}
\dot{g}(0)\in D \Rightarrow \dot{g}(t) \in D \hspace{3mm} \forall\; t\geq 0.
\end{align*}
Now, since $\mathbb{I}$ restricted to D is injective, on the cotangent side this is equivalent to ensuring that
\begin{align*}%
\mathbb{I}\dot{g}(0)\in D^* \Rightarrow \mathbb{I}\dot{g}(t)\in D^* \quad \forall\; t\geq 0.
\end{align*}
In other words,
\begin{align*}
P_{D^*_{\perp}}\mathbb{I}\dot{g}(0)=0\Rightarrow P_{D^*_{\perp}}\mathbb{I}\dot{g}(t)=0 \hspace{3mm} \forall \; t\geq 0.
\end{align*}
This is equivalent to
\begin{align*} \nabla_{\dot{g}}(P_{D^*_{\perp}}\mathbb{I}\dot{g}(t))=0.
\end{align*}
Recall that the covariant derivative ensures that, if a vector is of magnitude zero initially at a point on the curve, it remains so at all the other points on the curve if it is parallel transported along the curve.
So, equating this term in Equation \ref{LiebDperp} to 0, we get the expression for constraint force as
\begin{align}
\gamma_{\lambda}=-P_{D_{\perp}^*}\gamma-(\nabla_{\dot{g}}P_{D^*_{\perp}})\mathbb{I}\dot{g}. \label{cf}
\end{align}
This equation is as usual a generalization of the nonholomic system in the Euclidean space - the first term cancels the component of the force in the perpendicular direction and the second term is the centripetal term that arises due to the variation of the distribution $D$ itself from point-to-point, reflected by the expression $\nabla P_{D^*_{\perp}}$.
\begin{rem}
		For a constrained mechanical system, the constraint force calculated from  \eqref{cf} is naturally applied by the system on the body to enforce the physical constraint. But this has broader implications for controller design as well. In the design of controllers, these constraint force terms have to be necessarily incorporated in the controller dynamics to ensure that the controller dynamics respects the constraint. In other words, we mimic the constrained mechanical system in controller design to make the controller dynamics respect the given nonholonomic constraint. This will be further elaborated in the section on design of integral error.
\end{rem}
However, from the expression for the constraint force in  \eqref{cf}, it is not obvious that
$\gamma_{\lambda} \in D^*_{\perp}$ as required by Lagrange-d'Alembert principle
and through the physics. To verify this explicitly, we prove two Lemmas which will also be used to prove further results in controller design as well.
\begin{lemma}
	\label{projnlemma}
	Let $T^*G = D^* \oplus D^*_{\perp}$. Then
	\begin{align*}
	\mathbb{I}\dot{g}\in D^*\Rightarrow (\nabla_{\dot{g}}P_{D^*_{\perp}})\mathbb{I}\dot{g}\in D^*_{\perp}.
	\end{align*}
\end{lemma}
{\bf Proof}: As subspaces $D^*,D^*_{\perp}$ are complementary,
 \begin{align*}
P_{D^*}P_{D^*_{\perp}}\mathbb{I}\dot{g}=0.
  \end{align*}  Hence, it follows that:
  $\nabla_{\dot{g}}  (P_{D^*}P_{D^*_{\perp}}\mathbb{I}\dot{g} ) =0$. Applying the Leibnitz rule,
\begin{align*}
(\nabla_{\dot{g}} P_{D^*}) P_{D^*_{\perp}}\mathbb{I}\dot{g} + P_{D^*}(\nabla_{\dot{g}}P_{D^*_{\perp}})\mathbb{I}\dot{g} +
P_{D^*}P_{D^*_{\perp}}\mathbb{I}\nabla_{\dot{g}}\dot{g}=0
\end{align*}
In the above equation, the first term vanishes as we have assumed $\mathbb{I}\dot{g}\in D^*$ and the third term vanishes as it is $P_{D^*}P_{D^*_{\perp}}(.)$. So, the second term has to vanish and hence this completes the proof.
\qed
\begin{lemma}
	\label{cflemma}
	The constraint force $\gamma_{\lambda}=-P_{D_{\perp}^*}\gamma-(\nabla_{\dot{g}}P_{D^*_{\perp}})\mathbb{I}\dot{g} \in D_{\perp}^*$.
\end{lemma}
{\bf Proof:} The first term in  \eqref{cf} is in $D^*_{\perp}$ as it is projected into that subspace. The second term is also in $D^*_{\perp}$ and this follows from the previous Lemma.
\qed

\section{Generalizing the notion of error}
Let us now come to the main aim of generalizing the notion of error, derivative of error and integral of error - the three parts of PID control.
\subsection{Generalizing the error function - the Polar Morse Function}
A PID controller is uniquely characterised by three quantities of interest - namely the error, its derivative and its integral. Once we extend the notion of an error to a manifold, we substitute the covariant derivative instead of the usual derivative in Euclidean space to get the notions of derivative of error and integral of error respectively. The analogue of the error function $V$ that was $\frac{1}{2}||x||^2$ where $x$ is the configuration in Euclidean space (whose gradient gave us the error to use for proportional feedback) is now called a \textbf{polar Morse function}, which is defined below. The properties are inspired from the property of distance function in Euclidean space that are essential in guaranteeing convergence of the controller, the main idea being that the problem of regulating to a set point is reduced to minimizing the error function. Similarly, it is desirable to convert the stabilization problem to minimization of the polar Morse function $V$.
\begin{defn}
	A polar Morse function on a Lie group $G$ about a point $e\in G$ is a smooth real-valued function
	satisfying the following properties:
	\begin{enumerate}
		\item It has a unique global non-degenerate minimum at $e$ (whose value is without loss of generality assumed to be translated to 0) at the reference.
		\item It does not have local minima anywhere else.
		\item All its critical points are non-degenerate (so that the local behaviour of the function around critical points is completely characterised by the quadratic Hessian term in the Taylor expansion about the critical point).
	\end{enumerate}
\end{defn}
Hence, it is seen that, the problem of set point regulation to $e$, is same as  minimizing $V$ (since $e$ is the only global minimum of $V$). The fact that there are no local minima other than $e$ ensures that the trajectories do not get stuck at some other local minima other than $e$ in the process of reducing $V$. The non-degenerate nature of the critical points will help in establishing almost-global asymptotic stability.
The existence of such polar Morse functions was proven by Morse \cite{Morse}. We might also want the Morse function to not have any critical points at all  other than the global minimum (like the error function in Euclidean space). But this is typically not the case.  In compact manifolds, we cannot escape having critical points other than the reference point as any continuous function on a compact set achieves its maximum as well, and
hence there should be atleast one more critical point (the global maximum) apart from the global minimum at reference. So, if we have such a polar Morse function $V$, then, its differential $dV$ (denoted here as $\mathbb{I}grad(V)$) can be used as the notion of an error co-vector - a guide to what direction to go to increase $V$ as much as possible.
\begin{rem}
	If $G=\mathbb{R}^n$, observe that the function defined by $V=\frac{1}{2}||x-x_d||^2$ satisfies the above properties, its gradient $\nabla V=(x-x_d)$ is the familiar error vector in $\mathbb{R}^n$.
	The polar Morse function $V$ is a guide on $G$ as to what direction from a point leads to the maximum increase in error from the set point through its gradient.
\end{rem}
When we have a polar Morse function in a constrained system, with a distribution $D$, then the gradient of the Morse function $grad(V)$ in general will not lie in the distribution at all points. In fact it is difficult to analytically find such functions for a given distribution, even if its existence is guaranteed.

So, for a generic polar Morse function, one has to project its gradient to the distribution, to get a notion of error that is appropriate for the constrained system as only directions in the distribution are possible to access to. So even though travelling in the negative gradient is the best to minimize the error, it is not possible to do so, if the gradient vector is not in $D$ (note that the system velocity is constrained to lie in $D$). Hence the best thing to do in such a scenario as expected intuitively would be to project the gradient to $D$, that is, $P_D (grad V)$ in tangent space or $P_{D^*} dV$ in cotangent space should be the notion of error vector for a constrained system.

\subsection{Tracking to Regulation at $e$ - using group operation}
In Euclidean space, we can convert a trajectory tracking problem into a regulation problem about  a fixed origin, by defining the tracking error and restating the problem as that of making the tracking error to be zero.
Similarly,
we use the left translation to perform a similar trick of defining the tracking error in a Lie group. So, along the same lines as approached in \cite{DHSM}, given a reference trajectory $g_r(t)$, define the left invariant tracking error in a Lie group $G$ as %
\begin{align}
E(t)=(g_r(t))^{-1}\cdot g(t). \label{Edef}
\end{align}
On a Lie group,  $g(t) = g_r(t)$ is equivalent to making $E(t) = e$, the identity of the Lie group. Hence, the tracking problem has been converted into a regulation problem about $e$.

\begin{figure}[h]
	\center{\includegraphics[scale=0.5]{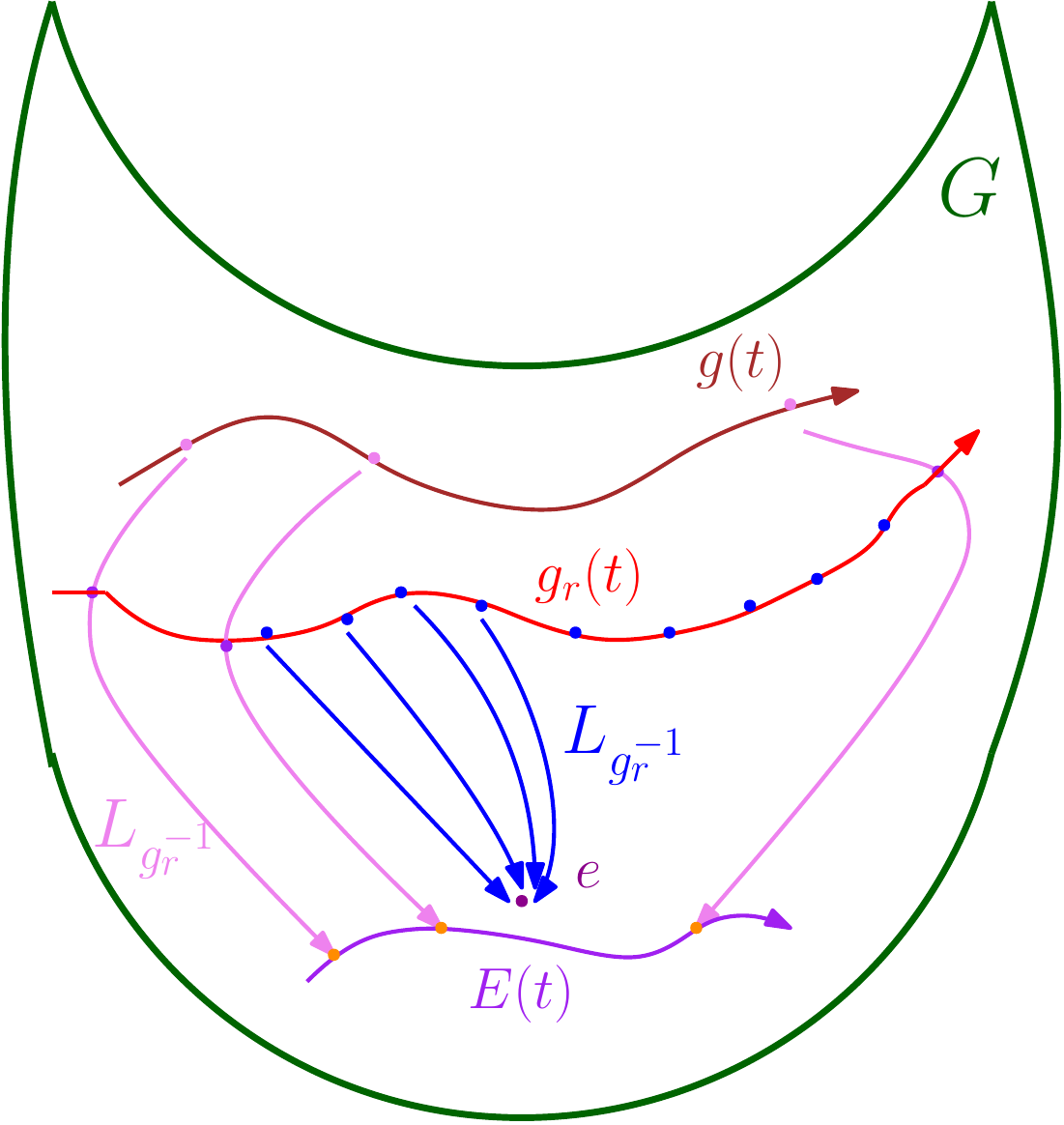}}
	\caption{Visualizing the error related terms}
	\label{fig:fig2}
\end{figure}

\begin{figure}[h]
	\center{\includegraphics[scale=0.5]{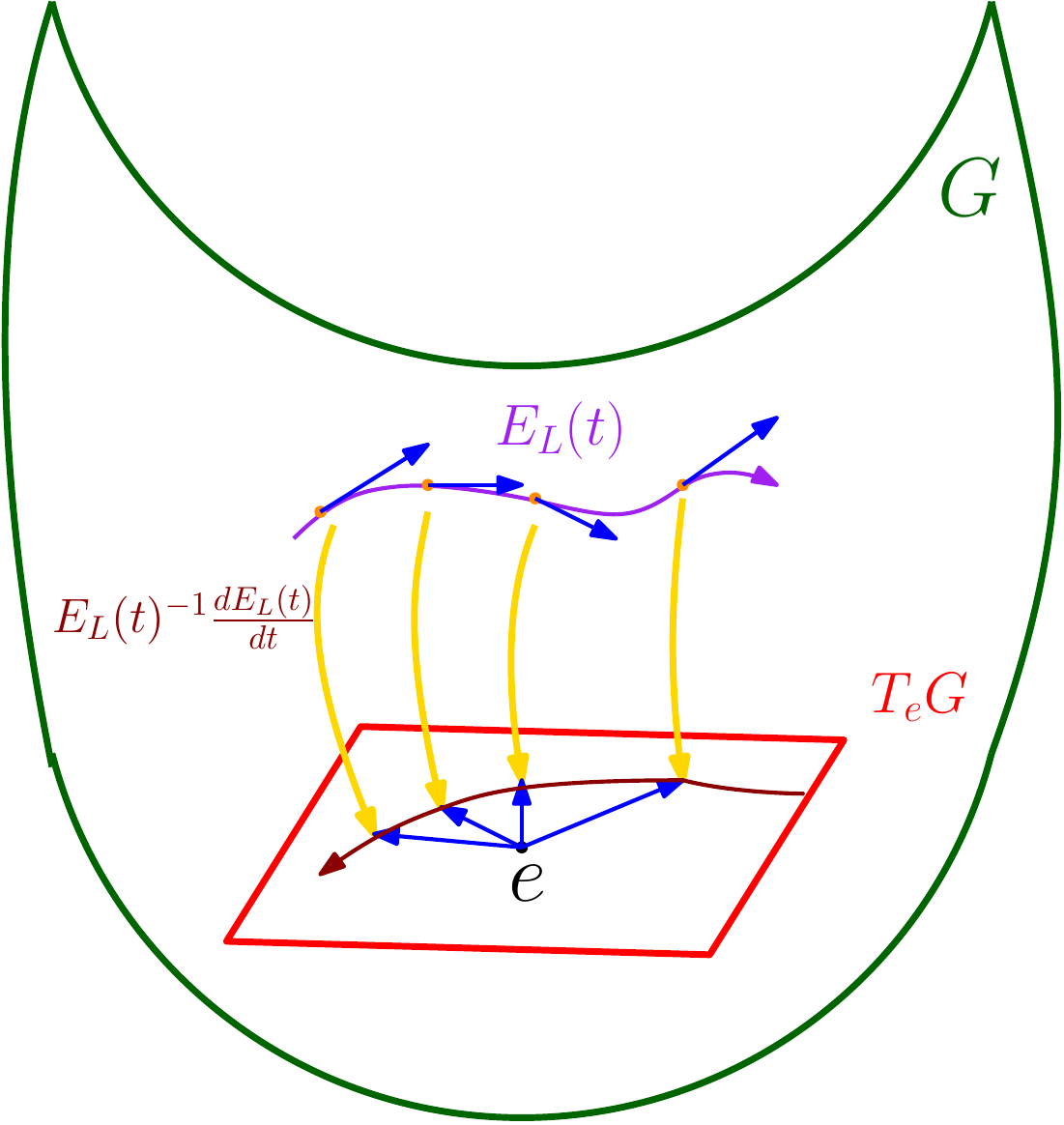}}
	\caption{Visualizing the error related terms}
	\label{fig:fig3}
\end{figure}

Having this, we can now construct a polar Morse function about the identity $e$, and define the appropriate tangent error vector. Once we define the left invariant tracking error, we can translate all the velocities (both for the actual curve and the reference curve trajectory) to the tangent space at identity (Lie algebra) through the derivative of the left translation map. (Note that for defining velocity of a curve, we do not need a covariant derivative).
Hence define the left velocity error (translated to the Lie algebra $\mathfrak{g}$) as
\begin{align}
 \zeta_E(t)=E^{-1}\cdot\dot{E}. \label{Velerrordef}
 \end{align}
  Also define left velocity as
  \begin{align}
  \zeta=g^{-1}(t)\cdot\dot{g}(t) \label{leftveldef}
  \end{align}and left velocity of reference trajectory as
  \begin{align}
   \zeta_r=(g_r(t))^{-1}\cdot\dot{g}_r. \label{leftrefveldef}
   \end{align}
    These are all visualized in Figures \ref{fig:fig2} and Fig \ref{fig:fig3}.
\begin{lemma}
	\label{productlemma}
	For any two smooth curves $\alpha(t)$ and $\beta(t)$ in G, we have
	\begin{align*}
		\frac{d}{dt}(\alpha(t) \cdot \beta(t))=\dot{\alpha}(t) \cdot \beta + \alpha \cdot \dot{\beta}(t).
	\end{align*}
\end{lemma}
Please refer  \cite{Frankel} for the proof.

Now, we need to generalize the notion of PID tracking control law for a Lie group.
It is generalization in the sense  that it reduces to the standard PID controller for $G=\mathbb{R}^n$ and $V=\frac{1}{2}||x-x_d||^2$, without constraints and whose treatments are analogous to the PID controller in  $\mathbb{R}^n$ wherein we exploit the group structure of an arbitrary Lie Group $G$ instead of the vector space structure for $\mathbb{R}^n$. For this, let us first take advantage of the fact that the metric is left invariant and the fact that every tangent vector can be translated naturally to the Lie algebra by left translations and every cotangent vector can be translated naturally to the dual space of the Lie Algebra by the pullback of the left translation map. And hence, any tensor in $T_gG$ can be translated to $\mathfrak{g}$.

More precisely, associate $X_g \in T_gG$ to the vector $g^{-1}.X_g$ in $T_eG=\mathfrak{g}$ and associate $\omega_g \in T^*_gG$ to $(L_g)^*\omega_g $ in $T^*_eG=\mathfrak{g}^*$.
A vector field $X$ is a smooth assignment of a tangent vector to every point in $G$. But now since every tangent vector $X_g$ can be translated to the Lie Algebra by the left translation (Use the map $g^{-1}X_g=(L_{g^{-1}})_*X_g$), this gives rise to a map $\xi_X : G \rightarrow \mathfrak{g}$ as $\xi_X(g) = g^{-1}.X(g)$ and similarly for co-vector fields (1 forms like force covectors) and any tensor fields as any tensor field can be translated to $\mathfrak{g}$ under the diffeomorphism generated by left translation.

However as the metric tensor is left invariant, it pulls back to a constant tensor on $\mathfrak{g}$, independent of g. So, $\mathbb{I}(g)=\mathbb{I}$ is a constant, independent of $g$.
So, the operation of covariant derivative as well can also be translated as a map $\xi_g \times \xi_g \rightarrow \xi_g$ where $\xi_g$ is the set of all smooth functions from $G$ to $\mathfrak{g}$.
That covariant derivative, for now denoted by $\nabla^L$ as $\nabla^L_{\xi_X(g)}\xi_Y(g):=g^{-1}\nabla_{X(g)}Y(g)$.
Similarly, the intrinsic Newton's equations (which are tensor equations ) can be translated and written on $\mathfrak{g}$. And the projection maps $P_D,P_{D_{\perp}},P_{D^*},P_{D^*_{\perp}}$ which are tensor fields can be translated as a smooth family of functions from $\mathfrak{g}\rightarrow \mathfrak{g}$ for each $g\in G$.

Henceforth, we will omit the superscript L or separate notations for objects pulled back to $\mathfrak{g}$ from $T_gG$. So the intrinsic Newton's equations, when pulled back on $\mathfrak{g}$, (with $\zeta$ defined as $g^{-1}\dot{g}$) we get
\begin{align}
P_{D^*}(\mathbb{I}\nabla_{\zeta}\zeta)&=P_{D^*}\gamma \label{pullbackmechG} \\
P_{D^*_{\perp}}(\mathbb{I}\nabla_{\zeta}\zeta)&=P_{D^*_{\perp}}\gamma + \gamma_{\lambda} \label{pullbackmechGperp}
\end{align}
with the $\nabla$ being understood as $\nabla^L$ and $\gamma$ as pull back of the control force to $\mathfrak{g}$.

Having done this, let us now determine the dynamics satisfied by the left invariant tracking error defined above, hereafter known as error dynamics as in standard control literature. We will derive an expression for the time evolution of the left velocity error $\zeta_E$ using the following Lemma, whose proof can be found in \cite{DHSM}.
\begin{lemma}
	\label{errorlemma}
	The error velocity $\zeta_E$ is related to the velocity of the trajectory by
	$\zeta_E=\zeta-Ad_{E^{-1}}\zeta_r$. The error dynamics is given by $\nabla_{\zeta_E}\zeta_E=\nabla_{\zeta}\zeta-F_r(E,\zeta_E,\eta_r)$ where
	$\eta_r \deff Ad_{E^{-1}}\zeta_r$ and $F_r(E,\zeta_E,\eta_r)=\nabla_{\zeta_E}\eta_r+\nabla_{\eta_r}\zeta_E+\nabla_{\eta_r}\eta_r$.
\end{lemma}
\begin{cor}
Note that when $g_r(t)$ is constant, as it in the case of regulation problem, then $\zeta_E=\zeta$ and $F_r=0$ as expected. $F_r$ is the analog of the feedforward term arising in trajectory tracking using classical PID control in $\mathbb{R}^n$.
\label{cor}
\end{cor}
In Euclidean space, the error in velocity $\zeta_E=\dot{\pmb{e}}$ (the difference between reference velocity and actual velocity compared after bringing them to the Lie algebra) and the rate of change of error in the direction of error velocity $\zeta$  which is $\nabla_{\zeta}grad(V)=\zeta$ are the same - which is $\dot{e}$. This compatibility depends on the metrical structure on $\mathbb{R}^n$ and the Morse function chosen specially as $\frac{1}{2}||\pmb{x}||^2$ being compatible with the metric. But one has to let this go when handling the generic case.

\begin{thm}
\label{estbdynsys}
    The dynamic equations for the trajectories of the mechanical system \eqref{pullbackmechG}-\eqref{pullbackmechGperp} evolving along the constraint distribution $D$, in the Lie Algebra $\mathfrak{g}$, is given by
    \begin{align*}
P_{D^*}(\mathbb{I}\nabla_{\zeta_E}\zeta_E)=P_{D^*}\gamma -P_{D^*}(\mathbb{I}F_r(E,\eta_r,\zeta_E)
.\end{align*}
\end{thm}
\textbf{Proof:} 
Now, for nonholonomically constrained systems, the error dynamics also can be split into one along the distribution and one perpendicular to it as
\begin{align}
P_{D^*}(\mathbb{I}\nabla_{\zeta_E}\zeta_E)&=P_{D^*}(\mathbb{I}\nabla_{\zeta}\zeta)-P_{D^*}(\mathbb{I}F_r(E,\eta_r,\zeta_E)) \label{errorD}  \\
P_{D^*_{\perp}}(\mathbb{I}\nabla_{\zeta_E}\zeta_E)&=P_{D^*_{\perp}}(\mathbb{I}\nabla_{\zeta}\zeta)-P_{D^*_{\perp}}(\mathbb{I}F_r(E,\eta_r,\zeta_E)) \label{errorDperp} .
\end{align}
Putting in the respective Newtonian expressions and the constraint forces derived  in  \eqref{cf}, \eqref{pullbackmechG}, and \eqref{pullbackmechGperp}, we get
\begin{align}
P_{D^*}(\mathbb{I}\nabla_{\zeta_E}\zeta_E)&=P_{D^*}\gamma -P_{D^*}(\mathbb{I}F_r(E,\eta_r,\zeta_E)) \label{errorfinalD}\\
P_{D^*_{\perp}}(\mathbb{I}\nabla_{\zeta_E}\zeta_E)&=-(\nabla_{\zeta}P_{D^*_{\perp}})\mathbb{I}\zeta-P_{D^*_{\perp}}(\mathbb{I}F_r(E,\eta_r,\zeta_E)) . \label{errorfinalDperp}
\end{align}
Hence, this proves the result. \qed

\section{PID control scheme}
Let us now motivate and define the PID controller for the dynamical system established in \eqref{estbdynsys} by considering each of the three terms separately.
\subsection{Proportional error - Definition and Compatibility with $D$}
We have seen already that the gradient of the polar Morse function should take the role of the error vector (which now is a tangent vector). But since we have a constraint now, and the control force is active only in the constraint subspace, we cannot use the actual gradient that may not belong to the constraint distribution $D$. So we have to project the gradient onto the constraint distribution and use it to define the proportional error vector - that is $P_D grad(V)$ (in the tangent space) or $P_{D^*}dV$ (in the cotangent space).  Since, it is not the case, in general, that the gradient of an arbitrary Morse function is in the distribution at all points, it has to be projected.
\subsection{Integral error - Definition and Compatibility with $D$}
Now we proceed with the PID tracking controller design. Before we do so, we define the notion of an integral error. In classical control theory on Euclidean spaces, the integral of error, denoted by $z(t)$, defined as $z(t)=\int e\hspace{1mm} dt$ serves as the definition of the integral error. It is also equivalently defined by the dynamical system $\dot{z}=e$. The right hand side of the equation is the error vector. But in a Lie group, one has to use the gradient of the Morse function which takes over the role of the error vector. Also, in a manifold, the covariant derivative takes the role of the derivative for vectors like integral error.

So, a natural candidate for the dynamics defining the integral error would be, replacing the derivative by the covariant derivative and error vector by the gradient of the Morse function in the Euclidean integrator, which is given as follows (as defined in \cite{DHSM})

\begin{align}
\mathbb{I}\nabla_{\zeta_E}\zeta_I=\mathbb{I}grad(V)=dV \label{intuncons}
\end{align}
where $\zeta_I$ is the integral error vector of the system. But this definition works fine for {\it unconstrained systems}. For system with constraints, when we
wish to incorporate such a term in the PID control, it is necessary that the error vector in the right hand side of the integral error dynamics is in the distribution.

This is achieved by taking the projected gradient instead of the actual gradient of the Morse function in the RHS. So a candidate definition would be
\begin{align}
    \label{tryint}
    \mathbb{I}\nabla_{\zeta_E}\zeta_I=\mathbb{I}P_Dgrad(V)=P_{D^*}dV.
\end{align}
But this alone does not ensure that $\zeta_I$ stays in $D$ once it is initialized in $D$. This is not ensured in  \eqref{tryint} by the term $P_Dgrad(V)$  lying in $D$ alone.
We wish to ensure that
\begin{align}
P_{D^*_{\perp}}\mathbb{I}\zeta_I(0)=0\Rightarrow P_{D^*_{\perp}}\mathbb{I}\zeta_I(t)=0 \hspace{3mm}  \forall\;  t\geq0 \hspace{3mm} \hbox{along $g(t)$.} \label{intreq}
\end{align}
In other words, when expressed in terms of the covariant derivative, we want
\begin{align}
 \nabla_{\zeta}(P_{D^*_{\perp}}\mathbb{I}\zeta_I)(t)=0 \;\;\; \forall\;  t\geq0. \label{intcovreq}
\end{align}
%
%
%
%
We will now first consider a \textit{constant set point regulation problem} when \textit{$\zeta_E=\zeta$}.
In this case, $\zeta_E$ can be put in \eqref{intcovreq} instead of $\zeta$ due to Corollary \ref{cor}.
Let us first examine what we get by taking the projection on the perpendicular space on both sides of  \eqref{intuncons}, assuming we have designed $V$ such that $\hbox{grad}(V)\in D$.  We then have
 \begin{align}
 P_{D^*_{\perp}}\nabla_{\zeta_E}\mathbb{I}\zeta_I=0. \label{intcovreqE}
\end{align}
However, this does not ensure that   \eqref{intcovreq} (with $\zeta$ instead of $\zeta_E$) would hold. Observe that the order of the projection and the covariant derivative are reversed in  \eqref{intcovreq} and \eqref{intcovreqE}.
So, it is necessary to introduce additional terms to the RHS of the integral error equation
defined in \eqref{intuncons}.
{\it Recall that this is exactly like adding a constraint force to ensure that $\nabla_{\dot{g}}P_{D^*_{\perp}}\mathbb{I}\dot{g}=0$ in the Newton's equation.}
So, defining the integral dynamics with the same inspiration as before, we propose
\begin{align}
\mathbb{I}\nabla_{\zeta_E}\zeta_I=P_{D^*}\mathbb{I}grad(V)+\gamma_I \label{trialdef}
\end{align}
where $\gamma_I$ is the, as yet unknown, constraint-force like term that we wish to add from $D^*_{\perp}$ to ensure that we get \eqref{intcovreqE}.
\begin{thm}
	\label{intdythm}
	Choosing $P_{D^*_{\perp}}\gamma_I=-(\nabla_{\zeta_E}P_{D^*_{\perp}})\mathbb{I}\zeta_I$ ensures
	the objective  $\nabla_{\zeta_E}(P_{D^*_{\perp}}\mathbb{I}\zeta_I)=0$.
\end{thm}
\textbf{Proof:}
As usual, projecting $\eqref{trialdef}$ along $D^*$ and along $D^*_{\perp}$ and writing its projected dynamics out on $D^*_{\perp}$, we get
\begin{align}
P_{D^*_{\perp}}(\mathbb{I}\nabla_{\zeta_E}\zeta_I)=P_{D^*_{\perp}}\gamma_I \label{perpintdylemmau}
\end{align}
as $P_{D}grad(V)\in D$.
So we have
\begin{align}
P_{D^*_{\perp}}(\mathbb{I}\nabla_{\zeta_E}\zeta_I)-P_{D^*_{\perp}}\gamma_I=0.\end{align}
Choosing
\begin{align}
P_{D^*_{\perp}}\gamma_I=-(\nabla_{\zeta_E}P_{D^*_{\perp}})\mathbb{I}\zeta_I. \label{gammaI}
\end{align}
 we have $\nabla_{\zeta_E}(P_{D^*_{\perp}}\mathbb{I}\zeta_I)=(\nabla_{\zeta_E} P_{D^*_{\perp}})\mathbb{I}\zeta_I+P_{D^*_{\perp}}(\nabla_{\zeta_E}\mathbb{I}\zeta_I)=0$. \qed

So we now define the integral error dynamics as
\begin{align}
\mathbb{I}\nabla_{\zeta_E}\zeta_I=P_{D^*}\mathbb{I}grad(V)-(\nabla_{\zeta_E}P_{D^*_{\perp}})\mathbb{I}\zeta_I. \label{interrorcons}
\end{align}
Notice how the above equation closely mirrors the Newtons's equation with $P_{D^*_{\perp}}\gamma_I$ taking the role of constraint force $\gamma_{\lambda}$ and ensures that the integrator dynamics, once initialized in $D^*$ stays in $D^*$ forever.
Invoking the same arguments as before, we can show by Lemma \ref{projnlemma} that the equation  $P_{D^*_{\perp}}\gamma_I=-(\nabla_{\zeta_E}P_{D^*_{\perp}})\mathbb{I}\zeta_I$ is consistent as the term $(\nabla_{\zeta_E}P_{D^*_{\perp}})\mathbb{I}\zeta_I$ is indeed in $D^*_{\perp}$. So indeed by Lemma \ref{projnlemma}, we can ensure that as in the constraint force case, $(\nabla_{\zeta_E}P_{D^*_{\perp}})\mathbb{I}\zeta_I$ is in $D^*_{\perp}$ if $\mathbb{I}\zeta_I$ is in $D^*$. So, it affects only the perpendicular component of the integral dynamics which is $P_{D^*_{\perp}}\mathbb{I}\nabla_{\zeta_E}\zeta_I$ and ensures that $\nabla_{\zeta_E}(P_{D^*_{\perp}}\mathbb{I}\zeta_I(t))=0$.
\subsection{Velocity error - Compatibility with $D$}
Since for regulation problem, $\zeta_E=\zeta$, the error velocity is in $D$ if the system velocity $\zeta$ is in $D$ which is ensured by the constraint forces.
\subsection{The PID control law}
Now we define the PID controller as
\begin{align}
P_{D^*}(\gamma)&=-k_pP_{D^*}\mathbb{I}grad(V)-k_d \mathbb{I}\zeta_E \nonumber\\
&- k_i\mathbb{I} \zeta_I+P_{D^*}(F_r(E,\eta_r,\zeta_E)) \label{PID}
.\end{align}
(Note that since we are dealing with regulation problem with constant set point, the feedforward term (the last term in \eqref{PID}) $P_{D^*}(F_r(E,\eta_r,\zeta_E))=0$).
We need to explicitly verify that the RHS of \eqref{PID} is indeed in $D^*$.
\begin{thm}
    The PID controller defined by \eqref{PID} is well-defined, that is, all its three terms belong to the co-distribution $D^*$. \label{welldefPID}
\end{thm}
\textbf{Proof:}
\begin{itemize}
\item The first term $P_{D^*}dV\in D^*$ by definition or if we use the gradient $dV$, then $dV$ can be be ensured to be in $D^*$ by using a special polar Morse function.  
\item The fact that $\mathbb{I}\zeta_I \in D^*$ was addressed before and the integrator dynamics was so defined
to ensure that $\mathbb{I}\zeta_I(t) \in D^* \; \forall \; t\geq 0$ when the integrator is initialized in $D^*$, that is $\mathbb{I}\zeta_I(0) \in D^*$. 
\item
 $\mathbb{I}\zeta_E \in D^*$, as mentioned previously follows from the fact that $\zeta \in D^*$ (the constraint force ensures it) for the constant setpoint case. So now we restrict to the constant set point case when $F_r$, the feedforward term also vanishes appropriately. \qed
\end{itemize}
So, we have that all the terms in the PID controller are in $D^*$ and hence the equation of the controller is consistent.
\section{Convergence analysis}
\subsection{Defining the closed-loop system}
For the closed-loop system (or controlled system) with state variables $G\times \mathfrak{g}\times \mathfrak{g}$ (configuration error, error velocity, integral error), we need to determine the error dynamics. Let us now analyze the behaviour of the controlled system.
\begin{thm}
\label{clt}
    If $\mathfrak{D}_g$ is the subspace associated with the distribution $D$  transported to the Lie algebra, therein denoted by $\mathfrak{D}_g$, then the closed-loop mechanical system of interest has a configuration space of $G\times \mathfrak{D}_g\times \mathfrak{D}_g$ for trajectories evolving in the distribution and its time evolution is given by
    \begin{align}
    \label{clrest}
    P_{D^*}(\mathbb{I}\nabla_{\zeta_E}\zeta_E)&=-k_pP_{D^*}dV-k_d \mathbb{I}\zeta_E - k_i\mathbb{I} \zeta_I\\
    \label{clrestt}
    P_{D^*}(\mathbb{I}\nabla_{\zeta_E}\zeta_I)&=P_{D^*}dV.
\end{align}
 The equilibria of the closed-loop system are $(g,\zeta_E,\zeta_I)=(g^*,0,0)$ where $g^*$ is any point of the polar Morse function $V$ satisfying $dV(g^*)=0$.
\end{thm}
\textbf{Proof:}
The closed-loop system, with the PID control law $P_{D^*}\gamma=-k_pP_{D^*} dV-k_d \mathbb{I}\zeta_E - k_i\mathbb{I} \zeta_I$ is given by
\begin{align}
P_{D^*}(\mathbb{I}\nabla_{\zeta_E}\zeta_E)&=-k_pP_{D^*}dV-k_d \mathbb{I}\zeta_E - k_i\mathbb{I} \zeta_I  \label{errdy}\\
P_{D^*_{\perp}}(\mathbb{I}\nabla_{\zeta_E}\zeta_E)&=-(\nabla_{\zeta}P_{D^*_{\perp}})\mathbb{I}\zeta-P_{D^*_{\perp}}(\mathbb{I}F_r(E,\eta_r,\zeta_E)). \label{perperrdy}
\end{align}
The integral law resolved along $D^*$ and $D^*_{\perp}$ gives
\begin{align}
P_{D^*}(\mathbb{I}\nabla_{\zeta_E}\zeta_I)&=P_{D^*}\mathbb{I}grad(V) \label{intdy}\\
P_{D^*_{\perp}}(\mathbb{I}\nabla_{\zeta_E}\zeta_I)&=-(\nabla_{\zeta_E}P_{D^*_{\perp}})\mathbb{I}\zeta_I. \label{perpintdy}
\end{align}
We are interested about the trajectories of the system that satisfy the constraint distribution. So it is not valid to treat the entire space $G\times \mathfrak{g}\times \mathfrak{g}$ as the state-space for the system as we are interested in evolution of only those trajectories that are in $D$ (the dynamics of $\zeta_E,\zeta_I$ ensure that once they start in $D$, they stay in $D$ forever). So let us consider the restricted dynamical system on $G\times \mathfrak{D}_g\times\mathfrak{D}_g$ where $\mathfrak{D}_g$ is the subspace of admissible velocities at $T_gG$, after pulling it back to the Lie Algebra $\mathfrak{g}$.  So, let us restrict our attention to the trajectories of the system in $G\times \mathfrak{D}_g\times\mathfrak{D}_g$.

The dynamics of the closed-loop system restricted to trajectories in $G\times \mathfrak{D}_g\times\mathfrak{D}_g$ is given by \eqref{clrest} and \eqref{clrestt}. So the equilibria for this system (with no other disturbances to it)  would be when the tangential parts of covariant derivatives vanish. So this happens when
\begin{align}
    \label{eq}
    P_{D^*}dV=0\\
    \zeta_E=0\\
    \zeta_I=0.
\end{align}
This proves the theorem. \qed

\subsection{Topological restrictions on stability to isolated equilibria}
Now, the equilibrium where we want all the trajectories to converge to is typically a single point, the special equilibrium chosen as $(e,0,0)$, the global minimum of $V$, for a typical regulation problem. However, for nonholomically constrained mechanical systems, there are topological restrictions that prevent existence of smooth, time-invariant control laws to guarantee even local asymptotic stability to a single equilibrium point. One standard example is Brockett's theorem as in \cite{Brockett}. Since the geometric PID controller is also smooth and time-invariant, this restriction applies. So, one can never, for such systems, guarantee local asymptotic stability to a single equilibrium point $(e,0,0)$ . This restriction must reflect on the Morse functions. Soon, it will be shown that the geometric PID controller guarantees global asymptotic stability of the trajectories to the set $C=\{(g,0,0)| P_{D^*}dV(g)=0\}$. In such a case, if the set $C$ is singleton, it implies asymptotic stability to the single equilibrium point which is forbidden by Brockett's theorem. So, one cannot design a Morse function for which set the $C$ is made up of a discrete set of points and hence $C$ has to be a non-zero dimensional submanifold of $G$. So, in nonholonomic systems where such restrictions apply, it is natural to deal with systems where the set $C$ is not made up of isolated points. In such systems, one has to then consider designing the PID controller such that asymptotic stabilization to a least non-zero dimensional submanifold of interest containing $(e,0,0)$, that is allowed topologically.
\subsection{D-critical points}
It is noted that the equilibrium of the system, the set $C$ is not just where the gradient vanishes but its projection along the distribution vanishes, that is, $P_{D^*}dV=0$. So, we now propose a series of definitions intended to capture the nature of such points and that which generalizes the second-order analysis of critical points to the constrained case.
\begin{defn}
        Let $V$ be a smooth function on a Riemannian manifold $G$ and let $D$ be a constant dimensional constraint distribution $D$. Then $x\in G$ is called a \textbf{D-critical point} of $V$ if $P_{D^*} dV\bigg|_x=0$.
        \end{defn}
In the language of the definition above, the equilibria of the closed-loop system are the D-critical points. In case of unconstrained systems, the set of critical points are typically isolated for non degenerate polar Morse functions but they are not so in case of nonholonomically constrained systems as was seen in the previous subsection. In that case, one has to look for controllers that stabilize locally asymptotically, not to the single point $e$ but to a lower dimensional submanifold of $G$ containing $e$. So one has to allow for the case when the set of all D critical points is a submanifold rather than a discrete set of points.
        \begin{defn}
            Let $V$ be a smooth function on $G$ with a distribution $D$. Then a \textbf{D-critical point} $x$ of $V$ is called \textbf{D- non degenerate} if the Hessian restricted to $D$ is non degenerate (as a matrix or a bilinear form) at $x$.
        \end{defn}
         \begin{defn}
       A smooth function $V$ is called \textbf{D-Morse} in $G$ if it has only D non-degenerate D-critical points.
\end{defn}
Note that in case of unconstrained systems, non-dengeneracy of critical points ensure they are isolated but it is not so, in the constrained case, for $D$ non-degenerate $D$-critical points as remarked previously.
\begin{defn}
Let $G$ be a Lie group with a left invariant metric. Let $D$ be a constraint distribution in $G$. A D-Morse function $V$ is called \textbf{D-polar Morse} function if it
\end{defn}
\begin{itemize}
    \item is $D$-Morse
    \item has a global minimum at $e$ (the value at the global minimum $e$ can be shifted to 0, without loss of generality, by adding an appropriate constant).
\end{itemize}
\subsection{The main result}
Let us now define a Lyapunov function to prove almost global asymptotic stability of the closed-loop mechanical system to the set of $D$- critical points, the set $C$. It is the same function as defined in \cite{DHSM} but now restricted to the closed-loop system evolving in the distribution.

Let $W:G\times \mathfrak{D_g}\times \mathfrak{D_g}\rightarrow \mathbb{R}$, chosen to be
 \begin{align}
W&=k_pV(E)+\frac{1}{2}\bigg<\zeta_E,\zeta_E\bigg>+\frac{\gamma}{2}\bigg<\zeta_I,\zeta_I\bigg>\nonumber\\
&+ \alpha\bigg<P_D grad(V),\zeta_E\bigg>+\beta\bigg<\zeta_I,\zeta_E\bigg> \nonumber \\  &+\sigma \bigg<\zeta_I, P_Dgrad(V)\bigg> \label{Wdef}
\end{align}
where $k_P,\alpha,\beta,\gamma,\sigma>0$ whose values will be determined in due course. Let $\chi$ be a compact subset of $G\times  \mathfrak{g}\times \mathfrak{g}$ and let $\chi_0=\chi \cap (G\times \mathfrak{D_g}\times\mathfrak{D_g})$ be the compact set containing $(e,0,0)$ in the restricted state-space. Notice that except the first term, all other terms are pure quadratic. To fix the first term not being quadratic (which will help in establishing its positive definiteness), and to also cope up with the fact that the error velocity $\zeta_E$ is not equal to $\nabla_{\zeta_E}grad(V)$, let us define the following quantities
\begin{align}
 \lambda&=\sup_{\chi_0}\frac{<P_Dgrad(V),P_Dgrad(V)>}{2V(E)}\\
\mu&=\sup_{\chi_0}\frac{<\zeta,\nabla_{\xi}P_Dgrad(V)>}{<\zeta,\xi>}
\end{align}
 to denote the upper bound on the Hessian. These quantities will help in bounding the Lyapunov function and its derivative, that will help us in establishing their positive or negative definiteness.
These are well-defined for unconstrained system when $\chi_0=\chi$ because the well-defined nature of the quantities inside the supremums used to define $\lambda,\mu$ as they follow from the fact that $V$ is a polar Morse function and the existence of the supremums follow from the fact that continuous functions achieve their supremum on a compact set (See \cite{Morse}). As far as constrained systems are concerned, this part needs to be formally verified. Proceeding further assuming $\lambda,\mu$ exists, we prove convergence results.
\begin{lemma}
	\label{Wlowerboundlemma}
$W$ is lower bounded from its global minimum by the following quadratic function below
	  \begin{align*}
	W&\geq\frac{k_P}{2\lambda}\bigg<P_Dgrad(V),P_Dgrad(V)\bigg>+\frac{1}{2}\bigg<\zeta_E,\zeta_E\bigg>\nonumber\\
&+\frac{\gamma}{2}\bigg<\zeta_I,\zeta_I\bigg>+	\alpha\bigg<P_Dgrad(V),\zeta_E\bigg>+\beta\bigg<\zeta_I,\zeta_E\bigg>\nonumber\\
&+\sigma \bigg<\zeta_I,P_D grad(V)\bigg>.
	\end{align*}
\end{lemma}

\textbf{Proof:} The first term in the inequality follows from the definition of $\lambda$. The other terms are same as in the definition of $W$. So, the inequality follows. \qed

\begin{lemma}
Denoting by $\dot{W}$, the derivative of the Lyapunov function $W$ along the trajectories of the closed-loop system, it is upper bounded by the following quadratic function through the inequality
\begin{align*}
    -\dot{W}&\geq\beta k_I \bigg<\zeta_I,\zeta_I\bigg>-(\gamma-\alpha k_I -\beta k_P)\bigg<\zeta_I,P_D grad(V)\bigg>\nonumber\\
    &-(k_I+\beta k_D-\sigma\mu)\bigg<\zeta_I,\zeta_E\bigg>\\&+(\alpha k_P-\sigma)\bigg<P_D grad(V),P_D grad(V)\bigg>\nonumber\\
    &-(\beta-\alpha k_D)\bigg<P_D grad(V),\zeta_E\bigg>\nonumber\\
    &+(k_D-\alpha \mu)\bigg<\zeta_E,\zeta_E\bigg>.
\end{align*}
\end{lemma}
\textbf{Proof:} First, computing $\dot{W}$ from the closed-loop dynamics, we have
\begin{align}
\dot{W}&=k_p\langle grad(V),\zeta_E \rangle + \langle \nabla_{\zeta_E}\zeta_E,\zeta_E \rangle + \gamma \langle \nabla_{\zeta_E}\zeta_I, \zeta_I \rangle\nonumber\\ 
&+ \alpha \langle \nabla_{\zeta_E}grad(V),\zeta_E \rangle +
\alpha \langle grad(V),\nabla_{\zeta_E}\zeta_E \rangle\nonumber\\
&+ \beta \langle \zeta_I,\nabla_{\zeta_E} \zeta_E \rangle  + \beta \langle \nabla_{\zeta_E}\zeta_I,\zeta_E \rangle + \sigma \langle \nabla_{\zeta_E}\zeta_I,grad(V) \rangle \nonumber\\
&+ \sigma \langle \zeta_I,\nabla_{\zeta_E}grad(V)\rangle. \label{wdot}
\end{align}
Substituting for the various covariant derivatives appearing in \eqref{wdot}, we get
\begin{align}
\dot{W}&= k_P \langle P_D grad(V),\zeta_E \rangle +  \langle -k_P  P_D grad(V)-k_d\zeta_E\nonumber \\
&-k_i \zeta_I \pmb{-P_{D^*_{\perp}}(F_r)-(\nabla_{\zeta}P_{D^*_{\perp}})\zeta},\zeta_E \rangle\nonumber\\
&+ \gamma  \langle P_D grad(V)-(\pmb{\nabla_{\zeta_E}P_{D^*_{\perp}})\zeta_I},\zeta_I\rangle \nonumber\\
&+\alpha  \langle P_D grad(V), -k_P P_D grad(V)-k_d\zeta_E-k_I \zeta_I \nonumber \\ &-\pmb{P_{D^*_{\perp}}(F_r)-(\nabla_{\zeta}P_{D^*_{\perp}})\zeta }\rangle \nonumber \\
&+\alpha \langle \nabla_{\zeta_E}P_D grad(V),\zeta_E \rangle\nonumber\\
& + \beta \langle \zeta_I, -k_p P_D grad(V)-k_d\zeta_E-k_i \zeta_I \nonumber \\ &-\pmb{P_{D^*_{\perp}}(F_r)-(\nabla_{\zeta}P_{D^*_{\perp}})\zeta} \rangle\nonumber\\
&+\beta \langle P_D grad(V)-\pmb{(\nabla_{\zeta_E}P_{D^*_{\perp}})\zeta_I  },\zeta_E \rangle\nonumber\\
&+\sigma \langle P_D grad(V)-\pmb{(\nabla_{\zeta_E}P_{D^*_{\perp}})\zeta_I},P_D grad(V) \rangle \nonumber\\ \label{wdotbigeqn}
&+\sigma \langle \zeta_I,\nabla_{\zeta_E}P_D grad(V) \rangle
.\end{align}
Except for the additional terms involving $(\nabla_{\zeta}P_{D^*_{\perp}})\zeta$, $P_{D^*_{\perp}}(F_r)$, $(\nabla_{\zeta_E}P_{D^*_{\perp}})\zeta_I$ that are shown in bold in \eqref{wdotbigeqn}, this expression is the same as that of the  unconstrained system derived in \cite{DHSM}. The terms shown in bold vanish on taking inner products with $\zeta_E,\zeta_I,P_Dgrad(V)$ due to Lemma \ref{projnlemma} and hence the expression for $\dot{W}$ is same as that of the unconstrained system, with the projected gradient replacing the gradient.

The feedforward term vanishes for regulation problem and hence the bold terms involving $F_r$ vanish.  The other terms in bold  are of the form $\bigg<X,\nabla_{\zeta}P_{D^*_{\perp}} Y\bigg>$ where $X,Y \in D$. But by Lemma \ref{projnlemma}, $\nabla_{\zeta}P_{D^*_{\perp}} Y \in D_{\perp}^*$ if $Y \in D$ and hence its inner product with a vector from $D$ vanishes. \qed

\begin{thm}
    \label{thmcldinv}
 Let $(G,\cdot,\mathbb{I},D)$ be a mechanical system. Let $g_r$ be a constant set point (regulation) and $E$, the left invariant error.  Let $V$ be a D-polar Morse function on $G$ let the system be subjected to PID control law with the gains chosen to satisfy
    	\begin{align*}
    	&0< \;k_D\\
	&0 < k_I<\; \frac{k_D^3(1-\delta^2)}{\mu} \\
	&\max\bigg[2\kappa k_D^2,\frac{\lambda k_I^2}{2k_D^4}\bigg(1+\sqrt{1+\frac{4k_D^3}{\lambda k_I^3}(k_I^2+4\kappa^2k_D^6)}\bigg)\bigg]< \;k_P
	\end{align*}
	where $0<\kappa<\frac{2}{\mu}$ and $\delta=\kappa \mu -1$.
	Then, the closed-loop system, starting on $\chi_0$, converges asymptotically to any one of the equilibrium points in $C$. Further, if $G$ is compact, $\chi_0$ can be chosen as $G\times \mathfrak{D}_g\times \mathfrak{D}_g$ and hence the domain-of-attraction is global.
\end{thm}
\textbf{Proof:} The proof of the theorem follows from the application of Lyapunov's stability theorem applied to the closed-loop system with the Lyapunov function $W$. Since the Lyapunov function $W$ is same as that of an unconstrained system (with projected gradient replacing the gradient), and $\dot{W}$ is also same as that of the unconstrained system (the extra constraint  forces terms in bold vanish in the expression of $\dot{W}$), the stability proof exactly mirrors the proof for the unconstrained case which is worked out in \cite{DHSM}.
\qed
\section{Example: The two wheeled mobile robot}
\begin{figure}[h]
	\center{\includegraphics[scale=0.8]{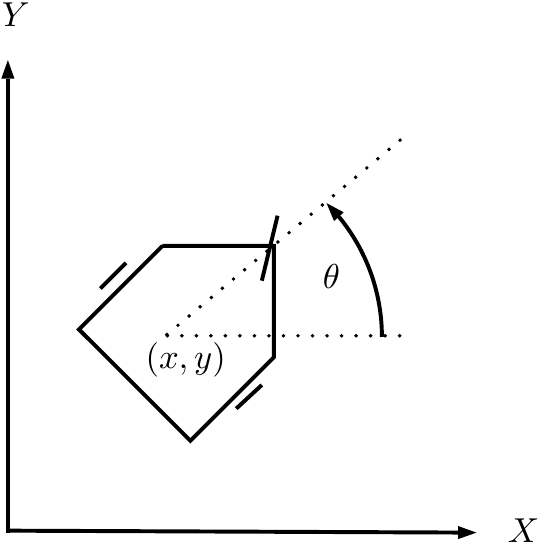}}
	\caption{Mobile robot}
	\label{fig:robot}
\end{figure}
In this section, the mobile robot, schematic shown in Figure \ref{fig:robot}, is taken as an example and the entire theory proposed in the article is worked out and demonstrated. The mobile robot has configuration space $G=\mathbb{R}^2\times\mathbb{S}^1=\{(x,y,\theta)| x,y\in\mathbb{R},\theta\in \mathbb{S}^1\}$. The first two coordinates are those of the coordinates of the centre of mass in the plane of the robot and the last circular coordinate is the orientation of it with respect to a body fixed frame. Note that as $\mathbb{R}^2,\mathbb{S}^1$ are separately Lie groups, with the operations of vector addition and angular displacement respectively, their direct Cartesian product $G$
is also a Lie group. With a local angular coordinates of the circle and assuming the robot to have a unit mass and unit moment-of-inertia of its rotational degree-of-freedom, its Lagrangian is given by
\begin{align}
    L=\frac{1}{2}(\dot{x}^2+\dot{y}^2+\dot{\theta}^2).
\end{align}
So we see that in these coordinates, the metric is locally Euclidean and hence $\mathbb{I}$ is identity in these coordinates. So, all the Christoffel symbols vanish and covariant derivatives reduce to derivatives of just their individual components.

But the static friction offered by the floor which ensures that the mobile robot rolls without slipping in the plane gives rise to a nonholonomic constraint on its velocities that in local coordinates $(\dot{x},\dot{y},\dot{\theta})$ is given as
\begin{align}
    \dot{x}\sin\theta-\dot{y}\cos\theta=0.
\end{align}
So, the constraint subspace is two dimensional (a single constraint eliminates one degree-of-freedom) and is verified easily to be spanned by the vectors
\begin{align*}
    D=\mathrm{span}\Big\{\begin{pmatrix}
    0\\0\\1
    \end{pmatrix},\begin{pmatrix}
    \cos\theta\\\sin\theta\\0
    \end{pmatrix}\Big\};
    D_{\perp}=\mathrm{span}\Big\{\underbrace{\begin{pmatrix}
    \sin\theta\\-\cos\theta\\0
    \end{pmatrix}}_{v}\Big\}.
\end{align*}
So, we have that all projection maps are matrices as the metric is locally Euclidean and hence we have
\begin{align}
    P_{D_{\perp}}=vv^\top=\begin{bmatrix}
    \sin^2\theta & -\sin\theta \cos\theta & 0 \\ -\sin\theta \cos\theta & \cos^2\theta & 0 \\ 0 & 0 & 0
        \end{bmatrix}.
\end{align}
Hence, the covariant derivative of $P_{D_{\perp}}$ is the ordinary time derivative of each of the entries of its matrix as the metric is locally Euclidean (Christoffel symbols vanish) and is given by
\begin{align}
    (\nabla P_{D_{\perp}}) = \begin{bmatrix}
    2 \dot{\theta}\sin\theta \cos\theta & (\sin^2\theta-\cos^2\theta)\dot{\theta} & 0 \\
     (\sin^2\theta-\cos^2\theta)\dot{\theta} & -2\dot{\theta}\sin\theta \cos\theta  & 0\\
     0 & 0 & 0
    \end{bmatrix}.
\end{align}
The constraint force is evaluated as
\begin{align}
    \gamma_{\lambda}&=-(\nabla P_{D^*_{\perp}})\mathbb{I}\dot{g}=- (\nabla P_{D^*_{\perp}})\begin{bmatrix}
    \dot{x}\\\dot{y}\\\dot{z}
    \end{bmatrix}\nonumber\\
    &=\begin{bmatrix}
    -2\dot{\theta}\dot{x} \sin\theta \cos\theta  +\dot{\theta} \dot{y}(\cos^2\theta-\sin^2\theta)\\ \dot{\theta}\dot{x}(\cos^2\theta-\sin^2\theta)+ 2\dot{\theta}\dot{y}\sin\theta \cos\theta \\0
    \end{bmatrix}.
\end{align}
So the constraint forces are as usual the centripetal and Coriolis like terms (quadratic in velocities as seen) that are responsible for keeping the trajectory of the particle inside the distribution.

Now, consider the control problem of stabilizing the mobile robot to a line with a fixed orientation on that line. As a concrete example, consider the problem of stabilizing the centre of the robot to the line $x=0$ (vertical y axis) with a horizontal orientation ($\theta=0$ or $\theta=\pi$). Consider the D-Morse function defined as
\begin{align*}
    V=\frac{1}{2}\bigg(x^2+y^2\bigg)+(1-\cos\theta).
\end{align*}
Its projected gradient is given by
\begin{align}
    P_Dgrad(V)=P_D\begin{bmatrix}
    x\\y\\\sin\theta
    \end{bmatrix}.
\end{align}
When evaluated explicitly by projecting into the span defined by its basis, it comes out to be
\begin{align}
\begin{bmatrix}
(x\cos\theta+y\sin\theta)\cos\theta\\(x\cos\theta+y\sin\theta)\sin\theta \\ \sin\theta
\end{bmatrix}.
\end{align}
One can verify that $\lambda=4,\mu=1$ for this function.
The set of $D$-critical points, which are the points at which the closed-loop system converges to, are the points where this projected gradient vanishes which is hence given by $\sin\theta=0,x\cos\theta+y\sin\theta=0$ which implies $\theta=0,\pi$ and $x=0$. So, the set of D-critical points is indeed the desired vertical axis with horizontal orientation.

\begin{figure}[h]
    \centering
    \subfloat[]{\includegraphics[width=0.5\textwidth]{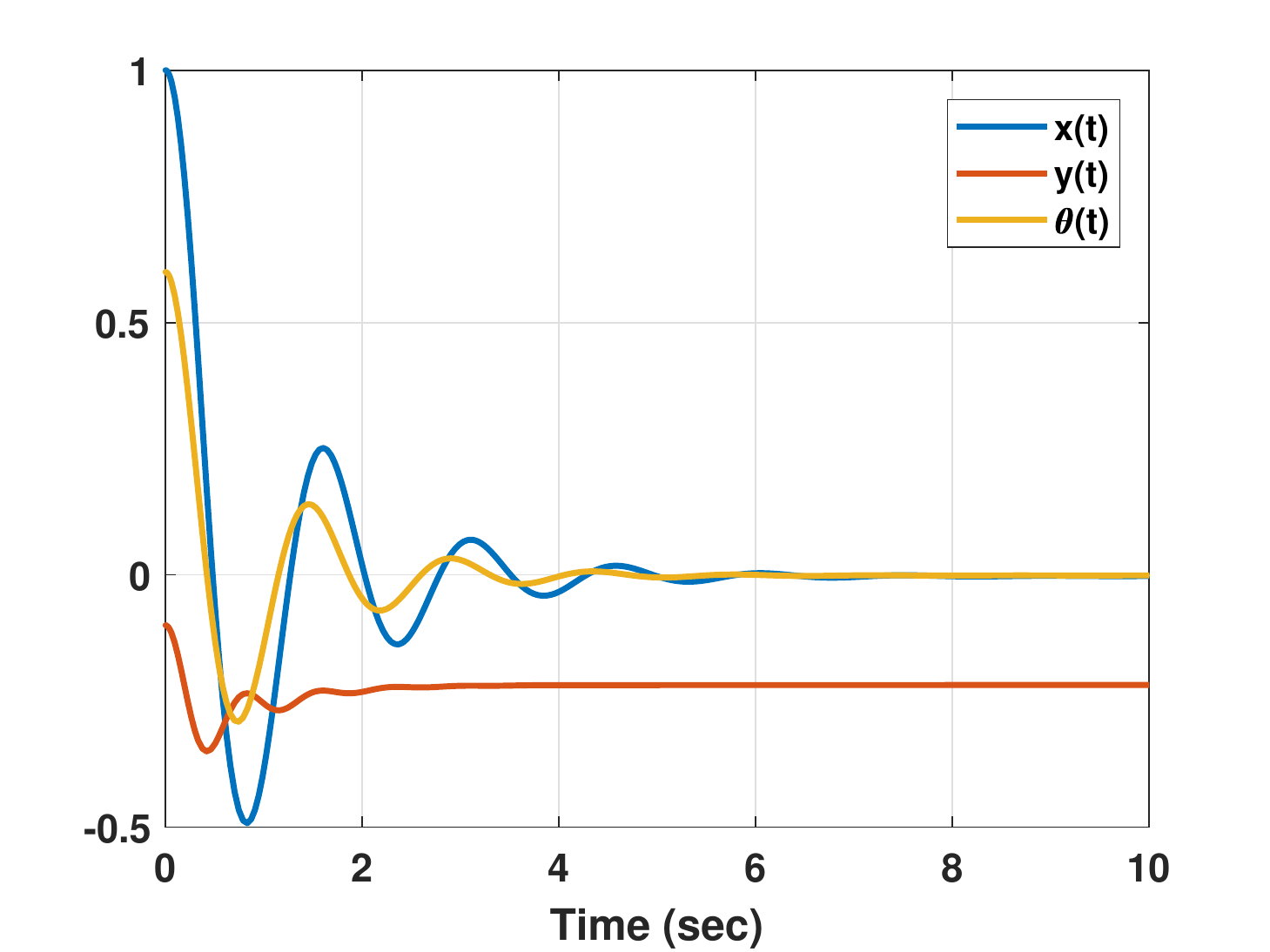}}\\
    \subfloat[]{\includegraphics[width=0.5\textwidth]{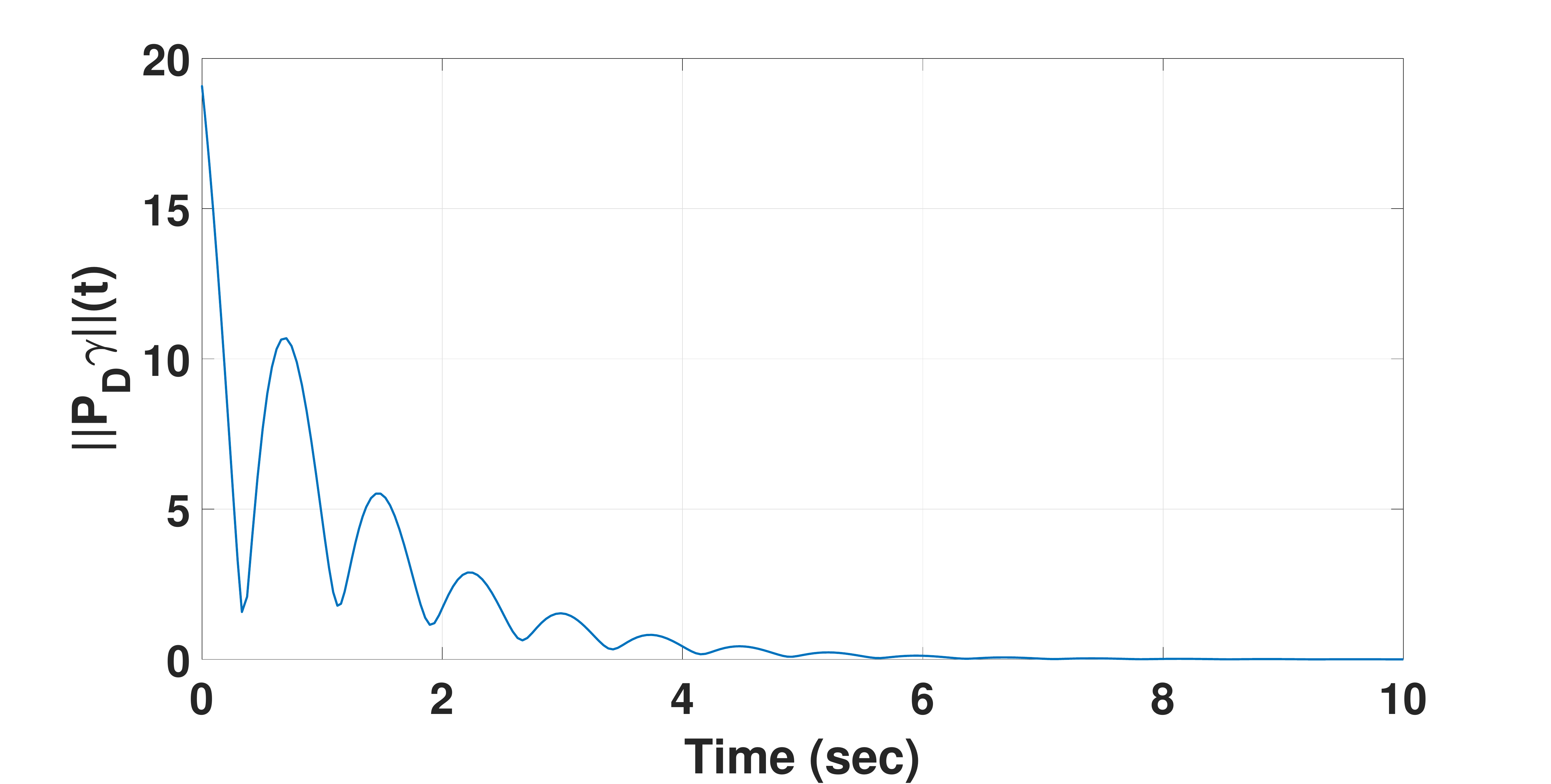}}
    \caption{Time response of states and magnitude of the control force for the mobile robot }
    \label{fig:robots}
\end{figure}

The simulation results are shown in Figure \ref{fig:robots} with $k_P=20,k_D=2,k_I=0.5$ for the initial condition chosen as $x(0)=1,y(0)=-0.1,\theta(0)=0.6,\dot{x}(0)=\dot{y}(0)=\dot{\theta}(0)=0$. 

\section{Conclusion and Future Work}
Thus, in this work, the design of a geometric PID controller to guarantee almost global asymptotic stability of a constrained mechanical system on a Lie group $(G,\cdot,D,\mathbb{I})$ has been carried out in the case of a constant desired set point or a desired submanifold. It is a general framework and can handle all classes of mechanical systems of interest - unconstrained systems in Lie groups, holonomically constrained systems in Lie groups (as in most cases, the integral submanifold of a holonomic constraint is not a Lie group), underactuated mechanical systems (with respect to the tangent space) become fully actuated in the constraint distribution $D$ (with respect to the constraint distribution) and hence become included in the present unified framework without any additional analysis.

Possible future work is indicated below.
\begin{enumerate}
    \item Work out practically important examples like spherical pendulum, Segway robot etc. and demonstrate the validity of the proposed controller computationally and in experiment.
    \item Extend the results to trajectory tracking scenario where the reference trajectory is time dependent. Here, it becomes challenging to characterize when and what trajectories can be tracked for a nonholonomically constrained system.
    \item Design an observer for the closed-loop system and analyze the coupled mechanical-controller-observer dynamics.
\end{enumerate}

\bibliographystyle{plain}      
\bibliography{nh_pid_arxiv}

\end{document}